\documentclass[twocolumn,showpacs,preprintnumbers,amsmath,amssymb,showkeys,nofootinbib,prd]{revtex4}

\usepackage{epsfig}
\usepackage[utf8]{inputenc}
\usepackage[T1]{fontenc}

\usepackage{graphicx}
\usepackage{dcolumn}
\usepackage{bm}
\usepackage{epstopdf}
\usepackage{color}
\usepackage{slashed}

\newcommand{\cb}{\overline c}

\begin{document}

\title{Quark-gluon vertex from the Landau gauge Curci-Ferrari model}

\author{Marcela Pel\'aez$^{a,b}$}

\author{Matthieu Tissier$^a$}

\author{Nicol\'as Wschebor$^b$}

\affiliation{\vspace{.2cm}
$^a$LPTMC, Laboratoire de Physique Th\'eorique de la Mati\`ere 
Condens\'ee, CNRS UMR 7600, Universit\'e Pierre et Marie Curie, \\
bo\^\i te 121, 4 
place Jussieu, 75252 Paris Cedex 05, France.\\
$^b$Instituto de F\'{\i}sica, Facultad de Ingenier\'{\i}a, Universidad de la Rep\'ublica,\\
J.~H. y Reissig 565, 11000 Montevideo, Uruguay.}

\date{\today}

\begin{abstract}
  We investigate the quark-gluon three-point correlation function
  within a one-loop computation performed in the Curci-Ferrari massive extension of
  the Faddeev-Popov gauge-fixed action. The mass term is used as a
  minimal way for taking into account the influence of the Gribov
  ambiguity. Our results, with renormalization-group improvement, are
  compared with lattice data. We show that the comparison is in
  general very satisfactory for the functions which are compatible
  with chiral symmetry, except for one. We argue that this may be due
  to large systematic errors {when extracting this function from} lattice simulations. The quantities
  which break chiral symmetry are more sensitive to the details of the
  renormalization scheme. We however manage to reproduce some of them
  with good precision. { The chosen parameters allow to
  simultaneously fit the quark mass function coming from the quark propagator
  with a reasonably agreement.}
  \end{abstract}

\pacs{12.38.-t, 12.38.Aw, 12.38.Bx,11.10.Kk}

\maketitle

\section{Introduction}
\label{sec_intro}

The precise shape of the quark-gluon vertex plays an essential role in
the understanding of many important properties of the infrared
behavior of QCD
\cite{Fischer:2003rp,Bhagwat:2004hn,Alkofer:2008tt}. This vertex is in
particular fundamental in the analysis of the origin of dynamical
chiral symmetry breaking in the context of truncated Dyson-Schwinger
(DS) equations. Indeed, the DS equation for the quark self-energy
requires the knowledge of this vertex, together with the gluon
two-point vertex function. Even if this kind of object depends on the
gauge-fixing, one expects that the gauge-invariant physical content of
the theory is encoded on its behavior. For example, the meson spectrum
can be studied by injecting such vertices and two-point functions in
appropriate Bethe-Salpeter equations, see for example
\cite{Matevosyan:2006bk,Karmanov:2008bx,Krassnigg:2009zh,Nicmorus:2010sd,Roberts:2011cf,Sanchis-Alepuz:2013iia,Fischer:2014xha}.
For these reasons, the determination of the
quark-gluon vertex has been a topic of intensive work in the past. It
has been first studied in the perturbative domain. A general
parametrization based on its symmetries and a one-loop calculation of
the abelian part in Feynman gauge was performed in
\cite{Ball:1980ay}. Some years later, the complete one loop
calculation in an arbitrary linear covariant gauge and in arbitrary
dimension was performed in \cite{Davydychev:2000rt} (where previous
partial results are reviewed also).  More recently the complete 2-loop
off-shell calculation has been performed in $d=4$
\cite{Gracey:2014mpa}. These perturbative studies lead to a
satisfactory description of this vertex at high momenta. However, the
infrared sector, which is of direct interest for the understanding of
chiral symmetry breaking, is not accessible to perturbation theory. To
cope with problem, mainly two methods have been used: lattice
simulations and DS equations.

In both of these approaches, the most convenient gauge is the Landau
gauge. From the lattice simulation side, it is well-adapted because
the gauge-fixing operation can be expressed in terms of an
extremization, which can be implemented efficiently on the
lattice. From the analytic side, this gauge has several
advantages. First, it is a linear gauge condition that manifestly
preserves Lorentz invariance. Second, among all linear covariant
gauges, it is the only one that includes the symmetries of the
Curci-Ferrari gauge \cite{Curci76,Delduc:1989uc,Tissier:2008nw}. This
implies, in particular, the existence of some non-renormalization
theorems, that simplify considerably the algebraic work
\cite{Taylor:1971ff,Dudal:2002pq,Gracey:2002yt,Wschebor:2007vh}. Third,
in this gauge, the gluon propagator is transverse. Consequently, only
the vertices where the gluon lines are multiplied by transverse
projectors contribute to the correlation functions. Accordingly, all
consideration below will be done in this gauge fixing.

There have been many analysis of the infrared behavior of the
quark-gluon vertex based in the study of truncated DS equations (for a
review, see \cite{Alkofer:2008tt}; for more recent work on the
subject, see \cite{Williams:2014iea,Aguilar:2014lha,Mitter:2014wpa}).  In parallel
the intermediate and infrared regime of the vertex has been studied by
lattice simulations.  Many tensorial components have been analyzed
\cite{Skullerud:2002ge,Skullerud:2003qu} for some particular
configuration of momenta. Moreover, for the tensorial component that
is already present at bare level, which is expected to be
``dominant'', more general configurations of momenta have been
considered \cite{Kizilersu:2006et}. 

 In continuum analytical approaches,
  the gauge fixing is generally implemented through the Faddeev-Popov
  construction, which is a perfectly valid procedure in the
ultraviolet regime, where standard perturbation theory works well and
where the Faddeev-Popov construction is well justified. However, for
momenta smaller or of the order of 1 GeV (which is the typical energy
scale of the strong interaction) this procedure is not fully
justified. The standard lore is that it becomes inappropriate because
the coupling grows and perturbation theory eventually becomes a very
bad approximation scheme. This argument must be tempered because the
real expansion parameter of QCD is of the order of $N \alpha_S/(4\pi)$
and because lattice simulations, at least in some renormalization
schemes, indicate that this parameter remains moderate (see, for
example \cite{Boucaud:2000ey}).  Accordingly it is reasonable to
believe that some sort of perturbation theory should give results that are in
qualitative agreement with experiments or lattice simulations in that
regime. There is, however, another reason why standard perturbation
theory should become inappropriate in the infrared regime: the
Faddeev-Popov procedure itself becomes questionable. This originates
in the existence of Gribov copies \cite{Gribov77} in covariant
gauge-fixings of non-abelian gauge theories that are ignored in the
Faddeev-Popov construction.  These copies play no significant role in
the ultraviolet regime but a systematic treatment of the associated
ambiguity remains an open problem in the infrared.

The most developed procedure which aims at taking into account the
effects of Gribov copies is probably the Gribov-Zwanziger (GZ)
construction \cite{Zwanziger:1989mf,Zwanziger:1992qr,Zwanziger:2001kw,Dudal:2008rm}. The
purpose of this approach is to restrict the functional integral to the
first Gribov region where the Faddeev-Popov operator is positive
definite. Gribov showed that in this region there are no copies that
are an infinitesimal transformation of one another. Unfortunately it
was shown later that there are {\em finite} gauge transformations that
preserve the Landau condition \cite{vanBaal:1991zw}. Consequently the
GZ procedure does not remove all copies. Moreover, the implementation of
the restriction of the functional integral to the first Gribov region
is not completely justified from first principles. 

Despite these difficulties, first studies of the GZ predicted that the
gluon propagator would vanish at zero momentum, instead of diverging
as in the bare theory. A similar behavior was found also as solutions
of truncated DS equations \cite{von Smekal:1997is,Alkofer:2000wg,Fischer:2003rp,Bloch03} and in Non-perturbative Renormalization
Group (NPRG) calculations \cite{Fischer:2004uk,Fischer08} (for a general review on this formalism, see \cite{Berges:2000ew}).
This kind of solution is called ``scaling solution''. The suppression
of the gluon propagator was rapidly confirmed by lattice simulations.
However, when lattice simulations became more precise they showed that
the gluon propagator in Landau gauge did not behave precisely as the
``scaling solution'' that DS had suggested, neither in quenched
lattice simulations
\cite{Sternbeck:2007ug,Cucchieri:2007rg,Cucchieri:2008fc,Sternbeck:2008mv,Cucchieri:2009zt,Bogolubsky:2009dc,Dudal:2010tf}
nor in the presence of dynamical quarks
\cite{Bowman:2004jm,Parappilly:2006si}. In fact, lattice simulations
favored a gluon propagator which tends to a finite value at low
momentum in $d=4$. This massive behavior (sometimes called
``decoupling'' solution) gives a propagator which is suppressed with
respect to the bare one but not as strongly as in the ``scaling
solution'' of DS equations. In parallel such decoupling solutions were
also obtained from DS and NPRG equations
\cite{Aguilar:2004sw,Boucaud06,Aguilar07,Aguilar08,Boucaud08,
  Fischer08,RodriguezQuintero10,Huber:2012kd}, as well as in a
``refined'' version of the GZ construction \cite{Dudal:2008rm}.

Nowadays, all approaches finally converge toward a massive-like gluon
propagator in Yang-Mills theory. Moreover, in what concerns the ghost
sector, there is also a consensus in favor of a massless behavior,
with a slight enhancement at low energy but with the tree level
power-law.  Finally, we note that DS and NPRG equations favor a
coupling constant bounded in the infrared at not too large values, see
for instance \cite{Huber:2012kd}.

All these observations led two of us to test wether a simple
perturbative analysis where a massive gluon propagator was introduced
at the bare level \cite{Curci76} could reproduce quantitatively the
lattice data for several correlation functions.  This lagrangian is
invariant under a modified Becchi-Rouet-Stora-Tuytin (BRST) symmetry
that ensures the renormalizability of the model
\cite{Becchi:1974md,Becchi:1975nq,Tyutin:1975qk}.  However, because of
the gluon mass, the associated transformation is not nilpotent.  As a
consequence, the textbook construction of a physical space based on
the co-homology of the BRST charge does not apply.  It must be
stressed that this difficulty is common to all methods that go beyond
the standard perturbative FP method. In particular, the standard
definition of the physical space does not apply to the GZ model but
even stronger, does not apply to the procedure implemented on
Landau-gauge lattice simulations. In fact the standard perturbative
definition based on the cohomology of the BRST operator includes in
the physical space the transverse gluons. But this is not satisfactory
for two reasons. First, on physical grounds, we know that gluons are
confined and are not part of the physical spectrum. Second, if
trasverse gluons were in the physical space, the spectral density
associated with their two-point function should be positive and
lattice simulations cleary show positivity violations
\cite{Cucchieri:2004mf,Bowman:2007du}.

All these considerations show that a proper definition of a physical
space is an open and important problem.  In this sense, the lack of a
nilpotent BRST charge in the Curci-Ferrari should not prevent us from
studying it in relation with the lattice simulations of correlation
functions. It is an open question to know if one can define a physical
space for the Curcci-Ferrari model where all states would have a
positive norm.

In the present article we exploit this
massive extension of the Faddeev-Popov Lagrangian in order to study
the quark-gluon vertex in Landau gauge.  Our aim is twofold.
First, we want to compare our findings with lattice
  simulation in order to have a further test of the ability of the
  Curci-Ferrari model to reproduce lattice simulations. As we show below, we are able to compute the {\it full}
vertex at one loop in any dimension, for all tensorial structures and
for any configuration of momenta. The comparison of this one loop result is in
reasonable agreement with all available lattice data (in many cases
the agreement is {very good}). In particular, we discuss in detail the
results of a certain scalar function ($\lambda_2$ in the terminology
of \cite{Skullerud:2002ge}) where a significant discrepancy has been
observed between lattice and DS results
\cite{Aguilar:2014lha,Williams:2014iea}. We discuss this discrepancy
in detail giving a very simple explanation of its origin.
Second (and
more importantly) as explained before, this vertex is very important
in various physical applications of DS and Bethe-Salpeter
equations. We expect that the present result will be useful in
future physical applications in these equations.

{ We want to stress here that, as it stands, the simple
  one-loop calculation presented in this article does not encode a
  spontaneous breaking of the chiral symmetry. However, the
  renormalization-group flow leads to a strong enhancement of the
  quark mass in the infrared, which mimicks, for all practical
  purposes, the behavior induced by this symmetry breaking. This is a
  key ingredient for reproducing the infrared behavior of the
  quark-gluon vertex. }

The rest of the article is organized as follows. In
Sec.~\ref{sec_model} we describe in more details our model and shortly
review the systematic comparison of one-loop computations with lattice
data. We give some details on the one-loop calculations in
Sec.~\ref{sec_one_loop} and finally present our results in
Sec.~\ref{sec_results}.

\section{The model}
\label{sec_model}

In this article, we pursue our systematic analysis of the correlation
functions of QCD in the Landau gauge, in a scheme where perturbative
calculations can be performed in a controlled way for all energies. As
explained in the introduction we consider the (euclidean)
Curci-Ferrari lagrangian \cite{Curci76}:
\begin{equation}
\label{eq_cf}
  \begin{split}
\mathcal{L}_A= &\frac{1}{4}F_{\mu\nu}^aF_{\mu\nu}^a +\partial_{\mu}\bar
c^a(D_{\mu}c)^a+ih^a\partial_{\mu}A_{\mu}^a  +\frac{m^2}{2}A_{\mu}^aA_{\mu}^a  
  \end{split}
\end{equation}
where $A_\mu^a$ is the gauge field, $c^a$ and $\cb^a$ are ghost and
antighost fields, $h^a$ is a Lagrange multiplier that enforces the
Landau gauge condition $\partial_\mu A_\mu^a=0$, $g$ is the coupling
constant and
\begin{align*}
F_{\mu\nu}^a&=\partial_{\mu}A_{\nu}^a-\partial_{\nu}A_{\mu}^a+gf^{abc}A_{\mu}^bA_{\nu}^c,\\
(D_{\mu}c)^a&=\partial_{\mu}c^a+gf^{abc}A_{\mu}^bc^c.
\end{align*}
{ It is important to recall that mass is introduced at the level of the
  gauge-fixed action. As a consequence, the bare propagators decrease at
  large momenta as $1/p^2$ and the mass induces a soft breaking of the
  BRST symmetry. As a consequence, all the ultraviolet properties of the
  standard Faddeev-Popov action are preserved to all orders in
  perturbation theory, in particular
  renormalizability and the behavior of correlation functions.}

One of the main consequences of the mass term is that the low energy
properties of the system are more regular than within the usual
Faddeev-Popov approach. In particular, it is possible to find
renormalization schemes in which the coupling constant remains finite
down to the deep infrared \cite{Tissier:2011ey}, at odds with the
findings of standard perturbation theory which presents a divergence
of the coupling constant at some energy of the order of 1~GeV. The
absence of such a Landau pole opens the way to a consistent treatment
of the low energy regime of the theory within perturbation theory.

The general idea was tested in \cite{Tissier:2010ts,Tissier:2011ey}
where the quenched gluon and ghost propagators were computed at
one-loop order. The comparison with lattice data is very satisfactory,
with a discrepancy of at most 10\% in the whole range of energy. The
analysis of the 3-point correlation functions (three gluons and
ghost-gluon) was performed under the same line of investigation in
\cite{Pelaez:2013cpa}. The comparison with lattice data is also very
good although the lattice data are more noisy in this case.

More recently, we have considered the influence of dynamical quarks on
the theory, which are governed by the Lagrangian:
\begin{equation}
  \label{eq_quarks}
  \mathcal{L}_\psi= \sum_{i=1}^{N_f} \bar\psi_i(-\gamma_\mu D_\mu + M_i)\psi_i,
\end{equation}
where $\gamma_\mu$ are euclidean Dirac matrices satisfying
$\{\gamma_\mu,\gamma_\nu\}=2\delta_{\mu,\nu}$, the flavor index $i$
runs over the $N_f$ quark flavors\footnote{In this article, we mainly
  consider degenerate quark masses. In this case, the sum over flavors
  can be replaced by $N_f$.} and the covariant derivative of the quark
field reads
\begin{equation}
  \label{eq_cov_dev}
D_{\mu}\psi=\partial_{\mu}\psi-igA_{\mu}^a t^a \psi.
\end{equation}
In \cite{Pelaez:2014mxa} we have included the one-loop contribution of
the quarks to the previously computed gluon propagator and have
computed the quark propagator to the same order. Again, the gluon
propagator compares very well with the lattice results.\footnote{Note
  that there is no direct influence of the quarks to the ghost
  propagator at one-loop. However, the presence of the dynamic quarks
  modify the $\beta$ functions, which gives an indirect effect on the
  ghost propagator. Again, the comparison of the ghost propagator with
  lattice data is very satisfactory.} The quark propagator can be
decomposed in two independent structures in spinor indices, often
called the mass (proportional to the identity in spinor indices) and
field renormalization (proportional to a Dirac matrix). The one-loop
results for the quark mass is in good agreement with the lattice
data. In particular, it reproduces the strong enhancement at energies
smaller than roughly 2 GeV, which is reminiscent of the chiral
symmetry breaking \footnote{ A similar behavior can be seen in the present
study in the quenched approximation, see Fig~\ref{fig_mass} below.}. However, we do
not expect that the present simple one-loop calculation can break
spontaneously the chiral symmetry.  On the other hand, the field
renormalization of the quark is not reproduced correctly, even at a
qualitative level. This can be attributed to the fact that the
one-loop contribution to this quantity is unusually small
\cite{Pelaez:2014mxa}. Indeed, when the gluon mass is set to zero, the
field renormalization has no contribution at one loop but it does have
contributions at two-loops, that, in consequence, become dominant.

This model has also been used in another line of research in a recent series of articles:
the extension of Yang-Mills theory at
finite temperature was considered within the same framework with the
goal of studying the deconfinement transition
\cite{Reinosa:2014ooa,Reinosa:2014zta,Reinosa:2015oua}. It was found that a simple
one-loop calculation can reproduce the known phenomenology [second order
transition for the SU(2) group and first order transition for
SU(3)]. A next-to leading calculation in the SU(2) case leads to a
quantitative agreement for the critical temperature. 

\section{One-loop calculation}
\label{sec_one_loop}

\begin{figure}[tbp]
  \centering
 \includegraphics[width=\linewidth]{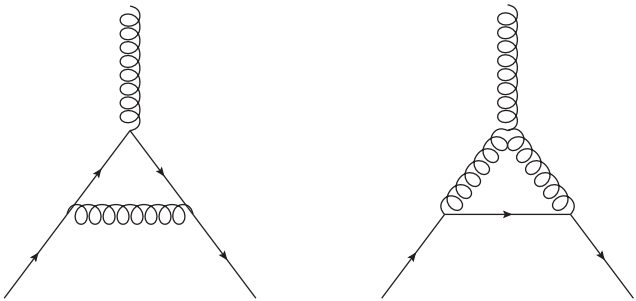}
\caption{One-loop Feynman diagrams for the quark-gluon vertex.}
\label{fig_diags} 
\end{figure}

The determination of the quark-gluon vertex at one loop requires to
compute the two diagrams depicted in Fig.~\ref{fig_diags}.
The result can be decomposed in twelve independent tensorial
structures. We follow here the convention of \cite{Skullerud:2002ge}
and express the quark-gluon vertex as
\begin{equation}
  \label{eq:decomp}
  \Gamma_{\bar\psi\psi A_\mu^a}(p,r,k)=t^a\Gamma_\mu(p,r,k)
\end{equation}
with
\begin{equation}
  \label{eq:decomp_2}
  \Gamma_\mu(p,r,k)=-ig\left(\sum_{i=1}^4 \lambda_i L_{i\mu}+ \sum_{i=1}^8 \tau_i T_{i\mu}\right).
\end{equation}
We give in Table \ref{tab_struc} the tensorial structures $L_{i\mu}$
and $T_{i\mu}$, together with some informations concerning the scalar
functions $\lambda_i$ and $\tau_i$.

\begin{table*}
\begin{tabular}{|c|l|c|c|}
\hline
Coupl. cte. & Tensorial structure & Dim. of cte&Small $M$\\
\hline
$\lambda_1$& $L_{1\mu}=\gamma_\mu$&(GeV)$^0$&$\mathcal O(M^0)$\\
$\lambda_2$& $L_{2\mu}=-(\slashed p-\slashed r)(p-r)_\mu$&(GeV)$^{-2}$&$\mathcal O(M^0)$\\
$\lambda_3$& $L_{3\mu}=-i(p-r)_\mu$&(GeV)$^{-1}$&$\mathcal O(M^1)$\\
$\lambda_4$& $L_{4\mu}=-i\sigma_{\mu\nu}(p-r)_\nu$&(GeV)$^{-1}$&$\mathcal O(M^1)$\\
\hline
$\tau_1$& $T_{1\mu}=i(k_\mu r_\nu k_\nu-r_\mu k^2)$&(GeV)$^{-3}$&$\mathcal O(M^1)$\\
$\tau_2$& $T_{2\mu}=(k_\mu r_\nu k_\nu-r_\mu k^2)(\slashed p-\slashed r)$&(GeV)$^{-4}$&$\mathcal O(M^0)$\\
$\tau_3$& $T_{3\mu}=\slashed k k_\mu-k^2\gamma_\mu$&(GeV)$^{-2}$&$\mathcal O(M^0)$\\
$\tau_4$& $T_{4\mu}=-i[k^2\sigma_{\mu\nu}(p-r)_\nu-2k_\mu\sigma_{\nu\lambda}r_\nu k_\lambda]$&(GeV)$^{-3}$&$\mathcal O(M^1)$\\
$\tau_5$& $T_{5\mu}=i\sigma_{\mu\nu}k_\nu$&(GeV)$^{-1}$&$\mathcal O(M^1)$\\
$\tau_6$& $T_{6\mu}=\slashed k(p-r)_\mu-k_\nu(p-r)_\nu\gamma_\mu$&(GeV)$^{-2}$&$\mathcal O(M^0)$\\
$\tau_7$& $T_{7\mu}=-\frac i2k_\lambda(p-r)_\lambda[(\slashed p-\slashed r)\gamma_\mu-(p-r)_\mu] -i (p-r)_\mu \sigma_{\nu\lambda}r_\nu k_\lambda$&(GeV)$^{-3}$&$\mathcal O(M^1)$\\
$\tau_8$& $T_{8\mu}=-\gamma_\mu\sigma_{\nu\lambda}r_\nu k_\lambda+r_\mu \slashed k-\slashed r k_\mu$&(GeV)$^{-2}$&$\mathcal O(M^0)$\\
\hline
\end{tabular}
\caption{We give the different tensorial structures on which the
  vertex was decomposed and the name of the associated (scalar)
  coupling constant. The last line describes the behavior of the
  coupling constant in the chiral limit. }
\label{tab_struc}    
\end{table*}

The calculations are performed by following the same procedure as
described in \cite{Pelaez:2013cpa,Pelaez:2014mxa}. The Feynman rules
are implemented in a Mathematica procedure and the integrals over
momentum written in terms of the Passarino-Veltman integrals
\cite{Passarino:1978jh}:
\begin{align}
  &A(m)=\int \frac {d^dq}{(2\pi)^d}\frac 1{q^2+m^2}\\
  &B_0(m_1,m_2)=\int \frac {d^dq}{(2\pi)^d}\frac 1{[q^2+m_1^2][(p+q)^2+m_2^2]}\\
  &C_0(p_1,p_2,m_1,m_2,m_3)=\\
&\nonumber\int \frac {d^dq}{(2\pi)^d}\frac 1{[q^2+m_1^2][(p_1+q)^2+m_2^2][(p_1+p_2+q)^2+m_3^2]}
\end{align}
Our expressions for the different
scalar functions expressed in terms of these integrals are available
in the supplemental material \cite{supplemental}. In actual
calculations, it is convenient to use the Feynman trick to perform the
integrals. In order to reduce the number of Feynman parameters, we
use the identity:
\begin{equation}
  \label{eq_ut}
  \begin{split}
&  \frac 1{(q^2+m_1^2)(q^2+m_2^2)}=\\
&\qquad\qquad\frac 1{m_1^2-m_2^2}\left(\frac 1{q^2+m_2^2}-\frac 1{q^2+m_1^2}\right)    
  \end{split}
\end{equation}
as many times as possible to reduce the number of propagators
appearing in the diagrams. The price to pay is that there appears
differences of masses in the denominator. For general momentum
configurations, after performing the integrals over the internal momentum, we are
left with a double integral over Feynman parameters. However, for
integer space dimension and for some momentum configurations that are
considered in next section, all integrals can be realized
analytically. We thus obtain analytic expressions for the one-loop
vertex for those configurations.

We have made several checks of our one-loop expressions. We have
compared our results with those of \cite{Davydychev:2000rt} which were
obtained in the limit of vanishing gluon mass. Our check consists in
comparing numerically our results for vanishing gluon mass with those
of \cite{Davydychev:2000rt}. This was done for several momentum
configurations and we find a perfect agreement. {
  Therefore, in the regime where the external momenta are large
  compared with the gluon mass, our results reproduce those of standard
  perturbation theory (including anomalous dimensions and all angular
  dependences). As a consequence, we do not study this regime further
  below.  }

Another check comes from the behavior of the coupling constants in the
chiral limit (limit of vanishing quark mass). In this limit, we easily
derive the Ward identity associated with the chiral symmetry:
\begin{equation}
  \label{eq_chiral_ward}
  \Gamma_{\mu}\gamma_5+\gamma_5\Gamma_{\mu}=0,
\end{equation}
where $\gamma_5=\gamma_0\gamma_1\gamma_2\gamma_3$. By using the
anticommutation relations of Dirac Matrices, it is easy to check that,
in the chiral limit, the structures $L_{3\mu}$, $L_{4\mu}$,
$T_{1\mu}$, $T_{4\mu}$, $T_{5\mu}$ and $T_{7\mu}$ are not compatible
with this Ward identity. We therefore conclude that the associated
coupling constants must vanish in the chiral limit. We have explicitly
verified that our one-loop expressions are compatible with this
result.

We regularize the theory by using the Infrared-Safe (IS)
scheme \cite{Tissier:2011ey}. As usual, we introduce renormalized
quantities, which are related to the bare ones by renormalization
factors:
\begin{align}
\nonumber  &A_B=\sqrt{Z_A}\, A,\qquad 
  c_B=\sqrt{Z_c}\, c,\qquad 
  \bar c_B=\sqrt{Z_c}\, \bar c,\\  &\psi_B=\sqrt{Z_\psi}\, \psi,\qquad 
  \bar \psi_B=\sqrt{Z_\psi}\, \bar \psi,\\
&g_B=Z_g g,\qquad m_B^2=Z_{m^2}m^2 ,\qquad M_B=Z_{M}M .
\end{align}
The different renormalization factors are defined by imposing:
\begin{align}
P^\perp_{\mu\nu}\Gamma_{A_\mu^a A_\nu^b}(p=\mu)&=(d-1)\delta^{ab}(\mu^2+m^2),\\
\Gamma_{c^a \cb^b}(p=\mu)&=\delta^{ab}\mu^2,\\
\Gamma_{\psi \bar \psi}(p=\mu)&=M-i\slashed p|_{p^2=\mu^2},\\
  Z_A Z_c Z_{m^2}&=1, \label{nonrenmcond}\\
  Z_g\sqrt{Z_A} Z_c&=1.
\end{align} 
The last two conditions are consequences of non-renormalization
theorems which relate the divergent parts of the different
renormalization factors
\cite{Taylor:1971ff,Dudal:2002pq,Gracey:2002yt,Wschebor:2007vh,Tissier:2008nw}. Here,
these relations are imposed also for the finite parts. The last
renormalization condition corresponds to the Taylor scheme and is
obtained by defining the coupling constant through the ghost-gluon
vertex function with vanishing momentum for the external ghost. The
condition (\ref{nonrenmcond}) can not be expressed directly in terms of vertex (or
correlation) functions. In particular, the mass parameter thus defined
is not directly related to the value of the gluon propagator at
vanishing momentum. Expressions for the renormalization factors can be
found in \cite{Tissier:2011ey,Pelaez:2014mxa}.

In the following, we use the renormalization-group improvement for the
quark-gluon vertex. It is obtained by integrating the RG equation for
the quark-gluon vertex function
\begin{equation}
\label{eq_rg_eq}
\begin{split}
\Big[ \mu \partial_\mu - \Big( \frac 1 2\gamma_A+ \gamma_\psi\Big)+\beta_g 
\partial_{g}+&
\beta_{m^2}\partial_{m^2}+\\&
N_f \beta_{M}\partial_{M}\Big]\Gamma_\mu=0,
\end{split}
\end{equation}
where
\begin{align*}
\beta_g(g,m^2,M)&=\mu\frac{dg}{d\mu}\Big|_{g_0, m^2_0,M_{0}},\\
\beta_{m^2}(g,m^2,M)&=\mu\frac{dm^2}{d\mu}\Big|_{g_0, m^2_0,M_{0}},\\
\gamma_A(g,m^2,M)&=\mu\frac{d\log Z_A}{d\mu}\Big|_{g_0, m^2_0,M_{0}},\\
\gamma_c(g,m^2,M)&=\mu\frac{d\log Z_c}{d\mu}\Big|_{g_0, m^2_0,M_{0}},\\
\beta_{M}(g,m^2,M)&=\mu\frac{dM}{d\mu}\Big|_{g_0, m^2_0,M_{0}},\\
\gamma_{\psi}(g,m^2,M)&=\mu\frac{d\log Z_{\psi}}{d\mu}\Big|_{g_0, m^2_0,M_{0}}.
\end{align*}
By integrating Eq.~(\ref{eq_rg_eq}) we relate the quark-gluon vertex
at different scales through the equation:
\begin{equation}
\label{eq_int_RG}
\begin{split}
\Gamma_\mu(\{p_i\},\mu_0,g(\mu_0),&m^2(\mu_0),M(\mu_0))=\\
&\frac{\Gamma_\mu(\{p_i\},\mu,g(\mu),m^2(\mu),M(\mu))}{\sqrt{z_A(\mu)}z_\psi(\mu)}.
\end{split} 
\end{equation} 
Here $g(\mu)$, $m^2(\mu)$ and $M_i(\mu)$ are obtained by integrating the
beta functions with initial conditions given at some scale $\mu_0$ and:  
\begin{equation*}
\label{eq_def_z_phi}
\begin{split}
\log z_A(\mu)&=\int_{\mu_0}^\mu\frac
     {d\mu'}{\mu'}\gamma_A\left(g(\mu'),m^2(\mu'),M(\mu')\right),\\ \log
     z_c(\mu)&=\int_{\mu_0}^\mu\frac
     {d\mu'}{\mu'}\gamma_c\left(g(\mu'),m^2(\mu'),M(\mu')\right),\\ \log
     z_\psi(\mu)&=\int_{\mu_0}^\mu\frac
     {d\mu'}{\mu'}\gamma_\psi\left(g(\mu'),m^2(\mu'),M(\mu')\right).
\end{split}
\end{equation*}
{ Observe that the running quark mass appears explicitely
  in the previous expressions. This leads to a back-reaction of this
  running mass on the vertex functions. As a consequence, the mass takes very different values at $\mu=0$ and
  $\mu=1$~GeV. Even at a perturbative level, this is a  large effect
  and, as discussed in the next section, is at the origin of the
  enhancement of some scalar components of the vertex.}

In the IS scheme, for the initial conditions of interest for comparing
with the lattice data, the coupling constant does not present a Landau
pole. In order to avoid large logarithms, we evaluate the
right-hand-side of Eq.~(\ref{eq_int_RG}) with $\mu$ of the order of
the external momenta {when they are larger than the mass.  In practice, we used
$\mu=\sqrt{(p^2+r^2+k^2)/2+m^2}$. It is important to stress that contrarily to what
happens in the pure glue sector, there is an significative dependence on the scheme
in the present calculation. In fact, it is important to chose the renormalization
point at a scale that remain fixed when all momenta go to zero as done in \cite{Tissier:2011ey}.}

{ In the following, we compare our results with lattice
  data which were obtained in the quenched
  approximation. Consequently, the beta functions for $g$ and $m$ as
  well as the anomalous dimensions for $A$ and $c$ are evaluated at
  $N_f=0$. Therefore, the quark mass does not back-react on these
  quantities, but it does back-react on the vertex.}

\section{Results}
\label{sec_results}

The results presented in the previous section were obtained for an
arbitrary momentum configuration and for all tensor structures. There
have been lattice studies of the quark gluon vertex for general
kinematics \cite{Kizilersu:2006et} for a particular tensorial
structure in the quenched approximation. Our results are { in general qualitative agreement}
with those of \cite{Kizilersu:2006et} at the level of precision of the
data. 

For some particular momentum configurations, lattice data were
obtained with larger statistics and therefore smaller error bars.  In
the following, we concentrate on these configurations which give more
stringent tests of the quality of our results. We perform a comparison
with lattice data in three momentum configurations. The first one
corresponds to configurations where the gluon has zero momentum with arbitrary
momentum of the quark. In this configuration many tensorial structures
were obtained in the quenched approximation \cite{Skullerud:2003qu}
and we compare below our results for all available functions.  The
second configuration corresponds to equal momenta for the quark and
antiquark (and, accordingly, gluon momentum equal to minus the double
of them). Again, many tensorial structures were obtained in the
quenched approximation for this configuration \cite{Skullerud:2003qu}
and we study all of them below. On top of these two families of
configurations, we consider a more general configuration with equal
momentum in modulus of quarks and anti-quarks but for various possible
angles between them (or, equivalently, different values of the modulus
of the gluon momentum). This configuration includes as particular
cases the previous ones but for general angles only a single tensorial
structure has been simulated \cite{Kizilersu:2006et} (also in the
quenched approximation). For the comparison with the simulations, we
focus on two values of modulus of the gluon momenta and also for the
completely symmetric configuration where the incoming momentum of the
gluon is equal in modulus to the quark and antiquark momenta.
 { It must be pointed out that the employed data are more
 that ten years old. More recent data with improved lattice parameter
 could allow to perform a more precise test of our curves but they do not exist for the moment.}

In the end of this section we finally discuss the effect of including dynamical
quarks for the present vertex. Even if there are no simulations of this
vertex in the unquenched case, there are some simulations of 2-point functions
\cite{Bowman:2005vx} that allows to fix the parameters
and, in consequence, we are able to make a prediction of the behaviour of
the quark-gluon vertex for the conditions that are present in those simulations.
These could be compared to future unquenched simulations of the quark-gluon vertex.

\subsection{Vanishing gluon momentum}
\label{vanishinggluon}
Let us first consider the case of zero gluon momentum. In this case,
 the tensor structure simplifies to the following form
\begin{equation}
\label{zerogluonmomentum}
\Gamma_\mu(p,-p,0)=
-ig\left[\lambda_1(p^2)\gamma_\mu-4\lambda_2(p^2)\slashed{p}p_\mu-2i\lambda_3p_\mu
\right].
\end{equation}
Moreover, all integrals appearing in the vertex functions can be
performed analytically. By introducing
\[\tilde{m}=\frac{m e^{\gamma/2}}{\sqrt{4\pi}},\]
\[\mathbb{L}_1=\log
   \left[\frac{\left(Y_1-p^2\right){}^2-\left(M^2-m^2\right)^2}{\left(Y_1+p^2\right){}^2-\left(M^2-m
   ^2\right)^2}\right],\]
\[\mathbb{L}_2=\log \left[\frac{\left(m^2-M^2-p^2\right)^2-Y_1^2}{\left(m^2-M^2+p^2\right)^2-Y_1^2}\right]\]and
\[Y_1=\sqrt{\left(m^2-M^2+p^2\right)^2+4 M^2 p^2},\]
we obtain in $d=4-\epsilon$
\begin{widetext}
\begin{align}
 \lambda_1(p^2)&=1+\frac{3Ng^2}{32\pi^2}\left[\frac{1}{\epsilon}-\log(\tilde{m})\right]\nonumber\\
&+\frac{g^2}{192 \pi ^2 m^4 N p^4} \log\left(\frac{M}{m}\right)   
\left\{-6 m^8 \left(N^2-1\right)+m^6 \left[M^2 \left(10 N^2-9\right)-\left(14 N^2+3\right) p^2\right]\right.\nonumber\\
&\left.-3 m^4
   \left[M^4 N^2-14 M^2 p^2+3 \left(N^2+2\right) p^4\right]+3 m^2 \left(M^6-7 M^4 p^2+15 M^2
   p^4-p^6\right)\right.\nonumber\\
&\left.-N^2 \left(M^2+p^2\right)^4\right\}-\frac{g^2}{192 \pi ^2 m^4 N p^4}\log \left(\frac{M^2+p^2}{M^2}\right)\left(M^2+p^2\right)^3  \left[N^2 \left(M^2+p^2\right)-3
   m^2\right]\nonumber\\
&+\frac{g^2}{384 \pi ^2 m^4 N p^4 Y_1}\left\{ m^2 p^2 Y_1 \left[12 m^4
    \left(N^2-1\right)+m^2 M^2 \left(6-8 N^2\right)+m^2 p^2\left(19 N^2+6\right)
   \right.\right.\nonumber\\
&\left.-6 \mathbb{L}_2 \left(2 m^2-5 M^2+p^2\right)
   \left(m^2-M^2+p^2\right)-2 N^2 \left(M^2+p^2\right)^2\right]\nonumber\\
&+\mathbb{L}_1 \left[m^4+2 m^2
   \left(p^2-M^2\right)+\left(M^2+p^2\right)^2\right] \left[6 m^6
   \left(N^2-1\right)+ m^4 M^2
   \left(3-4 N^2\right)\right.\nonumber\\
&+m^4\left.\left(8 N^2-3\right) p^2\right)\left.\left.-M^4m^2
   \left(N^2-3\right)+6m^2 M^2 p^2+m^2\left(N^2+3\right) p^4-N^2
   \left(M^2+p^2\right)^3
\right]\right\},
\end{align}
 \begin{align}
  \lambda_2(p^2)&=\frac{g^2}{384 \pi ^2 m^2 N p^4} \left[12 m^4 \left(N^2-1\right)-m^2 \left(4 N^2-3\right) \left(2 M^2-p^2\right)-2 N^2
   \left(M^2+p^2\right)^2\right]\nonumber\\
&-\frac{g^2}{384 \pi ^2 m^4 N p^6}\log\left(\frac{M}{m}\right)\left\{12 m^8 \left(N^2-1\right)-m^6 \left(10 N^2-9\right) \left(2 M^2-p^2\right)+6 m^4 \left(M^4
   N^2+p^4\right)\right.\nonumber\\
&+m^2 \left[-6 M^6-9 M^4 p^2+6 M^2 \left(5 N^2-1\right) p^4+9 p^6\right]\nonumber\\
&\left.+2 N^2
   \left(M^8+4 M^6 p^2-9 M^4 p^4+19 M^2 p^6+p^8\right)\right\}\nonumber\\
&-\frac{g^2}{384 \pi ^2 m^4 N p^6}\log \left(\frac{p^2}{M^2}+1\right)\left(M^2+p^2\right)^2  \left[m^2 \left(3 p^2-6 M^2\right)+2 N^2
   \left(M^2+p^2\right)^2\right]\nonumber\\
&+\frac{g^2}{768 \pi ^2 m^4 N p^8}\frac{\mathbb{L}_1}{Y_1}\left\{12 m^{10} \left(N^2-1\right) p^2+m^8 p^2 \left[M^2 \left(30-32 N^2\right)+\left(22 N^2-21\right)
   p^2\right]\right.\nonumber\\
&+m^6 p^2 \left[2 M^4 \left(13 N^2-9\right)+3 M^2 \left(1-2 N^2\right) p^2+\left(10
   N^2-9\right) p^4\right]\nonumber\\
&-m^4 \left[150 M^8 N^2+6 M^6 \left(N^2+1\right) p^2+M^4 \left(10
   N^2-3\right) p^4+2 M^2 \left(2 N^2-3\right) p^6+3 p^8\right]\nonumber\\
&+m^2 \left[300 M^{10} N^2+2 M^8
   \left(3-149 N^2\right) p^2+M^6 \left(4 N^2+15\right) p^4+9 M^4 p^6-M^2 \left(4 N^2+3\right)
   p^8\right.\nonumber\\
&\left.\left.-\left(2 N^2+3\right) p^{10}\right]+150 M^8 N^2 Y_1^2-2 N^2 \left(M^2+p^2\right)^2 \left(75
   M^8+M^6 p^2+3 M^4 p^4+3 M^2 p^6+p^8\right)\right\}\nonumber\\
&-\frac{g^2}{128 \pi ^2 m^4 N p^2}\mathbb{L}_2 \left(m^2+5 M^2 N^2\right) \left(m^2-M^2+p^2\right),
 \end{align}
and
 \begin{align}
  \lambda_3(p^2)&=\frac{g^2 M}{64 \pi ^2 m^4 N p^4}\left\{2 \log \left(\frac{M}{m}\right) \left[3 m^4 \left(N^2-1\right) (m^2-M^2) - N^2 p^4\left(m^2-4
   M^2\right)-p^4m^2\right.\right.\nonumber\\
&\left.-p^2 (m^2-M^2)  \left(m^2-3 M^2 N^2\right)-N^2 p^6\right]-6 m^4
   \left(N^2-1\right) p^2\nonumber\\
&-\frac{3 \mathbb{L}_1}{Y_1} m^4 \left(N^2-1\right) \left[p^2
   \left(m^2+M^2\right)+\left(m^2-M^2\right)^2\right]\nonumber\\
&-p^2 \mathbb{L}_2  \left(m^2-M^2+p^2\right) \left[m^2+N^2 \left(p^2-3 M^2\right)\right]\left.\right\}.
 \end{align}
 \end{widetext}

 When performing lattice simulations, these functions are obtained
 from the simulated vertex $\Gamma_\mu(p,-p,0)$ by projecting in the
 various tensor structures in the following way
 \cite{Skullerud:2002ge,Skullerud:2003qu}:
\[\lambda_1(p^2)=\frac{-1}{4g_B}\text{Im}\left(\text{Tr}\,\gamma_\mu
  \Gamma_\mu(p,-p,0)\Big|_{p_\mu=0,\ p_\nu\neq 0\ \text{for }\mu\neq\nu } \right),\]
\begin{equation}
\label{eqL2}\lambda_2(p^2)=\frac{1}{4p^2}\sum_\mu
\left(\frac{1}{4g_B}\text{Im}\left[\text{Tr}\, \gamma_\mu\Gamma_\mu(p,-p,0)\right]+\lambda_1(p^2)\right)
\end{equation}
and
\[\lambda_3(p^2)=\frac{1}{2p^2}\sum_\mu p_\mu
\frac{1}{4g_B}\text{Re}\left[\text{Tr}\, \Gamma_\mu(p,-p,0)\right]\]
where, in these expressions, no implicit sum over repeated indices is
meant. It is interesting to note that, in order to extract the
function $\lambda_2(p^2)$, a contribution of $\lambda_1(p^2)$ must be
added. This issue will be discussed in more detail below.

\begin{figure}[tbp]
  \centering
  \includegraphics[width=.9\linewidth]{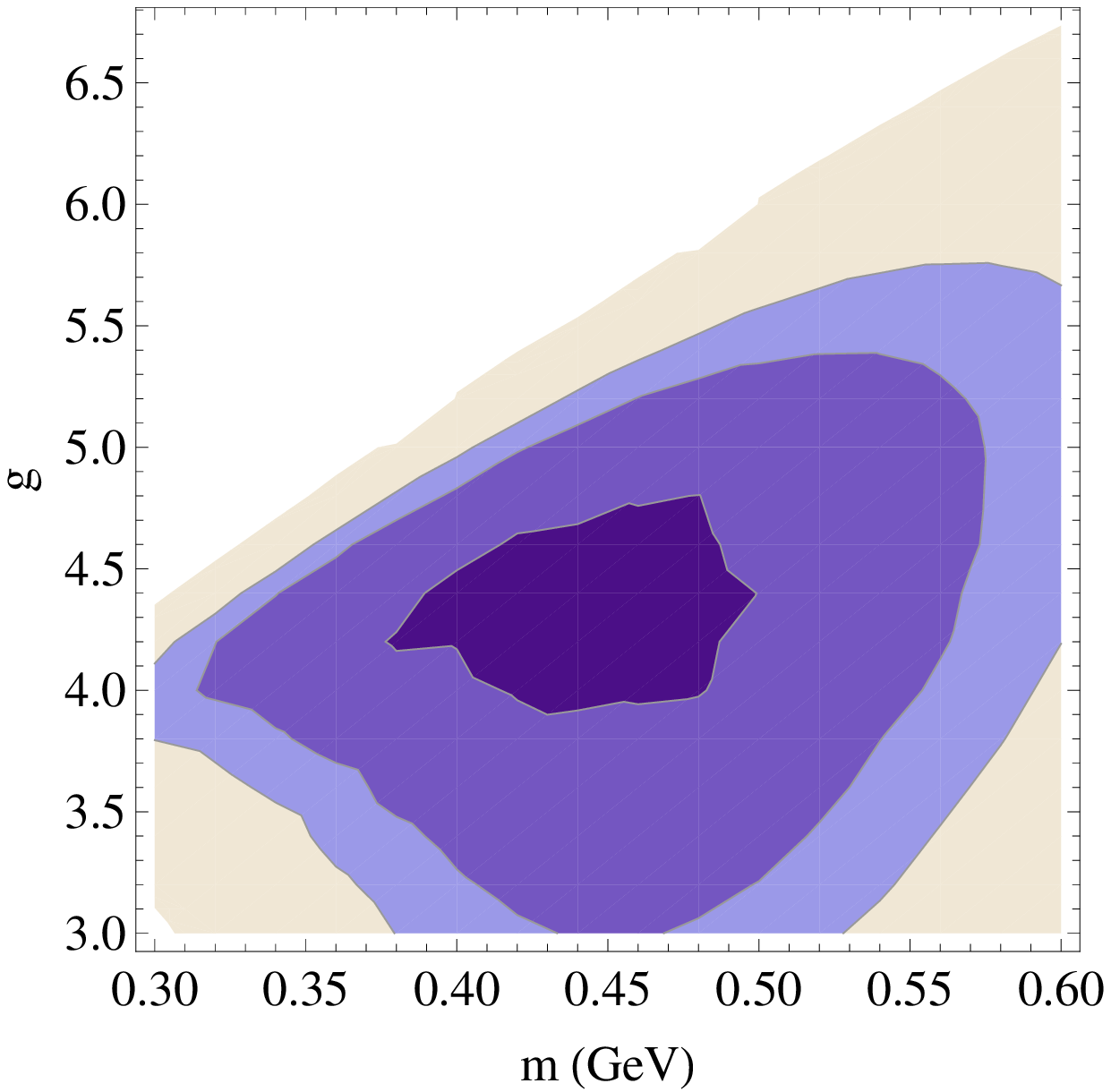}
  \includegraphics[width=.9\linewidth]{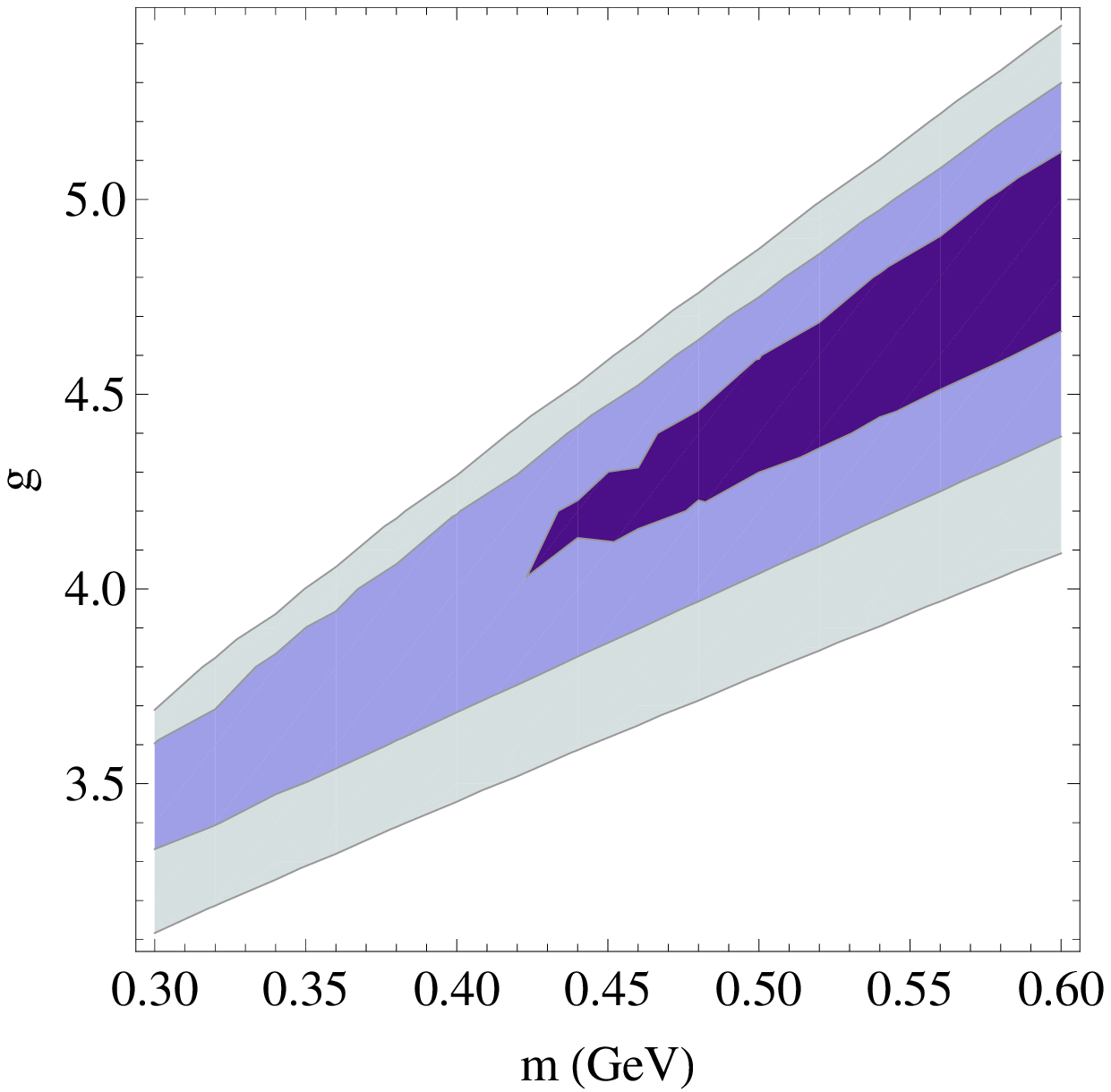}
  \caption{Contour levels for the error functions $\chi_{AA}$ for the
    gluon propagator (upper pannel, for 19\%, 22\% and 24\%) and
    $\chi_{c\cb}$ for the ghost propagator (lower
    pannel, for 4\%, 7\% and 10\%).}
  \label{fig_error_AACC}
\end{figure}

To compare our results with the lattice simulation of
\cite{Skullerud:2003qu} we renormalize the model in the IS scheme and
implement the corresponding renormalization group as discussed in
Sect.~\ref{sec_one_loop} . It is important to stress that the
simulations of \cite{Skullerud:2003qu} have been done in the quenched
approximation. Accordingly, we use the beta-functions of the pure
Yang-Mills theory derived in \cite{Tissier:2010ts,Tissier:2011ey} for
the gluon mass and coupling constant. This implies that the coupling
constant and the gluon mass can be fixed once the gluon and ghost
2-point functions have been fitted. The only remaining free parameter
is the valence quark mass (and the overall normalization factor). To
determine the best fit parameters for the ghost and gluon propagators,
we consider the errors functions:
\begin{align}
\label{eq_chi}
\chi^2_{AA}&=\frac{1}{4N}\sum_i(\Gamma_{\rm
lt.}^\perp(\mu_0)^2+\Gamma_{\rm
lt.}^\perp(p_i)^2)\left(\frac{1}{\Gamma_{\rm
lt.}^\perp(p_i)}-\frac 1{\Gamma_{\rm th.}^\perp(p_i)}\right)^2\nonumber\\
\chi^2_{c\cb}&=\frac{1}{4N}\sum_i( J^{-2}_{\rm lt.}(\mu_0)+
J^{-2}_{\rm lt.}(p_i))
\left(J_{\rm lt.}(p_i)-J_{\rm th.}(p_i)\right)^2\nonumber\\
\end{align} 
We represent these functions in Fig.~\ref{fig_error_AACC}.
The best fits are obtained for  $g=4.2$
and mass $m=0.44$~GeV at the scale 1 GeV which compare well with those
found in \cite{Tissier:2011ey}. As in that reference, this value has
been compared to other values of the coupling by taking into account the
RG running, finding a good agreement with best estimates up to a 10 \% error. 

\begin{figure}[tbp]
\centering
 \includegraphics[width=\linewidth]{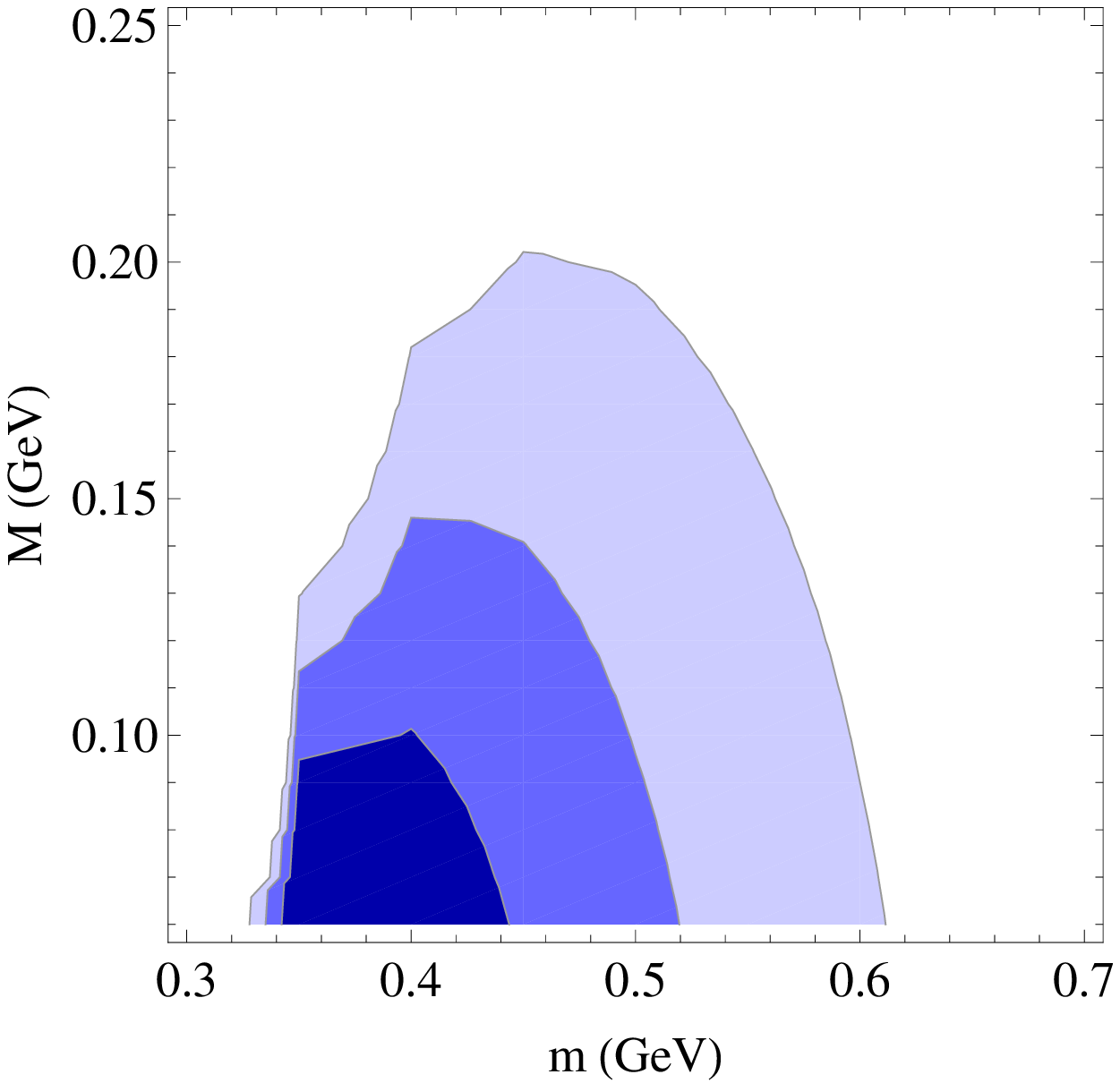}
 \includegraphics[width=\linewidth]{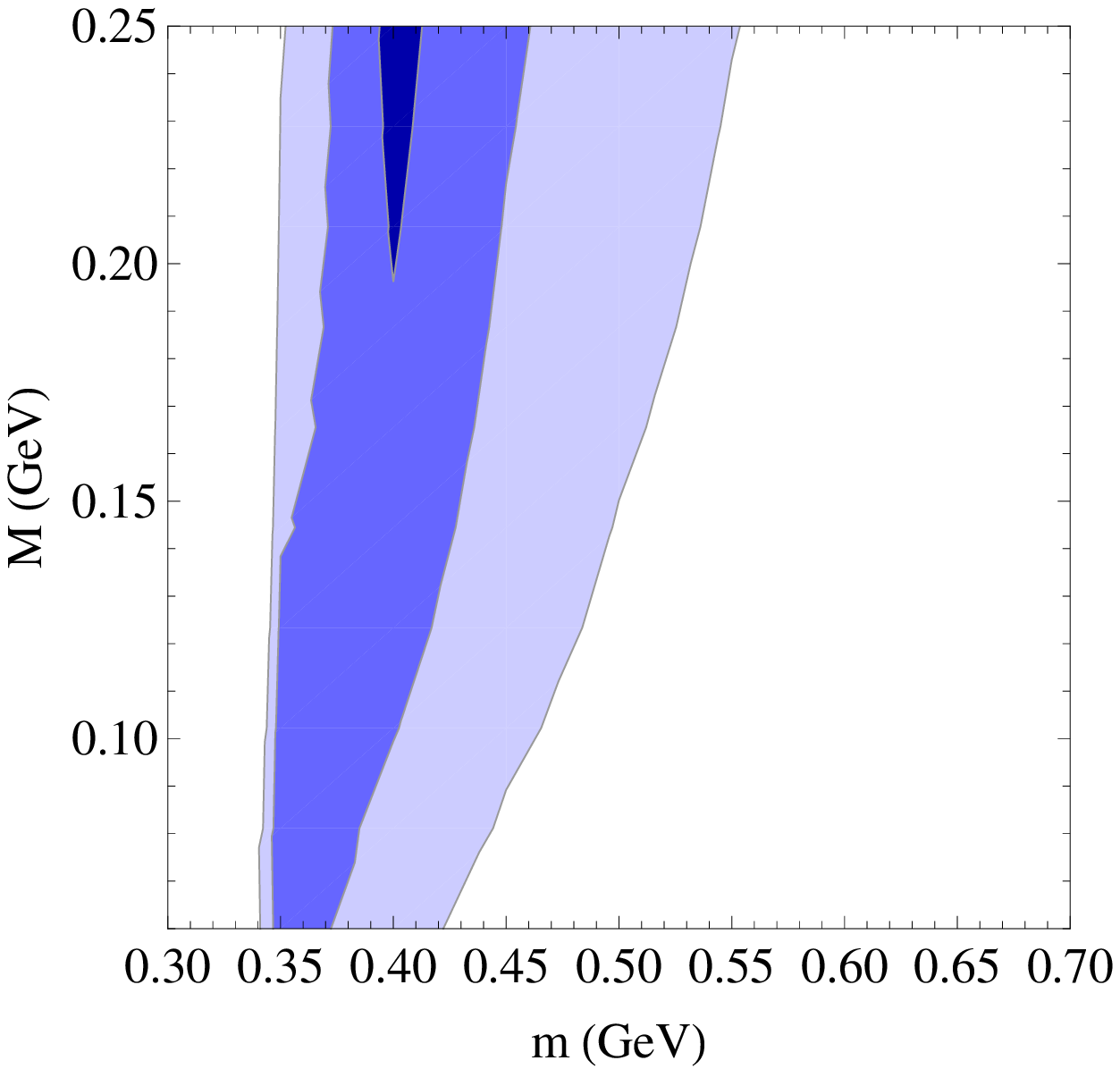}
\caption{\label{Errorl1} Contour levels for the quantity
  $\chi_{\lambda_1}$ (upper panel) and $\chi_{\lambda_3}$ (lower
  panel) in the plane of gluon mass vs. quark mass.
  The contour lines correspond to $4\%$, $4.5\%$ and $5\%$ for $\chi_{\lambda_1}$ and
  $10\%$, $20\%$ and $30 \%$ for $\chi_{\lambda_3}$.} 
\end{figure}

\begin{figure}[ht]
  \centering
  \includegraphics[width=.9\linewidth]{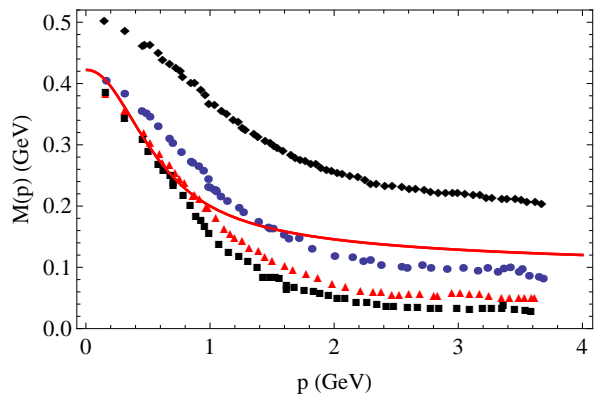}
  \caption{Quark mass as a function of the momentum. The
    renormalization group low is initialized at 1 GeV with $g=4.2$,
    $m=0.44$~GeV, and $M=0.2$~GeV. Lattice data \cite{Bowman:2005vx} are marqued
    with dots and parametrized by their bare masses: $M_0=19$~MeV (squares), $M_0=37$~MeV (triangles), $M_0=75$~MeV (circles),
$M_0=187$~MeV (diamond).}
  \label{fig_mass}
\end{figure}

{ The valence quark mass has been fixed in order to fit properly the
quenched quark propagator. More precisely we choose the quark mass in order
for the constituent quark mass to match the lattice value. Doing so, one obtains a
mass function (the scalar part of the quark propagator, see \cite{Pelaez:2014mxa}) that
reasonably agree with the lattice data from \cite{Bowman:2005vx} as can be seen in Fig~\ref{fig_mass}.
One see that this curve mimick the enhancement of the mass in the infrared
  but not the spontaneous chiral symmetry breaking. Indeed, the
  constituent quark mass vanishes when the current mass at 1 GeV tends
  to zero.
The corresponding
value of the quark mass is $M=0.2$~GeV at the scale $\mu= 1 GeV$, and it gives
a constituent quark mass $M(p=0)=0.42$~GeV. These values tend to overestimate the mass
function $M(p)$ in the UV by approximatively 30\% (for example, at 3 GeV we obtain
0.13 GeV instead of 0.10 GeV) and to underestimate the propagator
in the IR (below 1 GeV) by a maximum value of approximatively 18\%. Having said that,
the overall behavior is approximatively correct. It must be pointed out when fixing
the parameters in order to compare both to the quark mass function and to the
vertex functions, a practical difficulty takes place. The simulations have been
done for both quantities with different mass parameters. For the quark mass function
many bare masses were employed. In the vertex case, the employed values for the bare valence quark mass
were $M_0=115$~MeV and $M_0=60$~MeV  \cite{Skullerud:2003qu}. However the authors claim that the vertex functions are
almost insensitive to the bare quark mass and employ the data corresponding to $M_0=115$~MeV. In order to have a single set of data we choose the
bare valence quark nearest to this value that has been used for propagators
which is $M_0=75$~MeV. This introduces a systematic error but that do not seem
to be large because for relatively small masses like those \footnote{These values for the mass are not small with respect
to actual physical values of the masses for light quarks. However, they are relatively small with respect to typical QCD scales and, in
consequence, some quantities are not very sensitive to them.}, the mass function do not
seem to depend on $M_0$ too much (see \cite{Bowman:2005vx} and Fig.~\ref{fig_mass}).

One can also analyze the dependence on the valence quark mass of the vertex by
studying the error level for the function
$\lambda_1(p^2)$ defined as
\[\chi^2_{\lambda_1}=\frac{1}{N}\sum_i \frac 1{[\lambda_{1,\rm
    lt.}(\mu_0^2)]^2}\left[\lambda_{1,\rm lt.}(p_i^2)-\lambda_{1,\rm
    th.}(p_i^2)\right]^2,\]
We also consider the analogous error level
for the function $\lambda_3(p^2)$. For this last function the total error may be
misleading as discussed below because it average some momenta where the the curve
is essentially within error bars with the lowest momenta where errors are much larger.}
\footnote{As discussed in detail
  below, the comparison of our results with lattice simulations for
  the function $\lambda_2(p^2)$ is problematic. Consequently, we do
  not use this function for fitting the parameters.}  They are both
shown in Fig.~\ref{Errorl1}.
{ We observe that, even if the optimum value of the masses do not
coincide when fitting $\lambda_1(p^2)$ and $\lambda_3(p^2)$, there is
a region of parameters where both errors are reasonably small (that is
less than $10\%$). Moreover, the value of the gluon mass extracted
from the ghost and gluon propagators ($m=0.44$~GeV) is within an
error region with less than $20\%$.}  It is important to stress that
once the parameters $g$, $m$ and $M$ have been fixed, they are used
for the calculation of the quark-gluon vertex for all momentum
configurations. Accordingly, they become pure predictions without free
parameters of the present calculation.

\begin{figure}[tbp]
\centering
 \includegraphics[width=\linewidth]{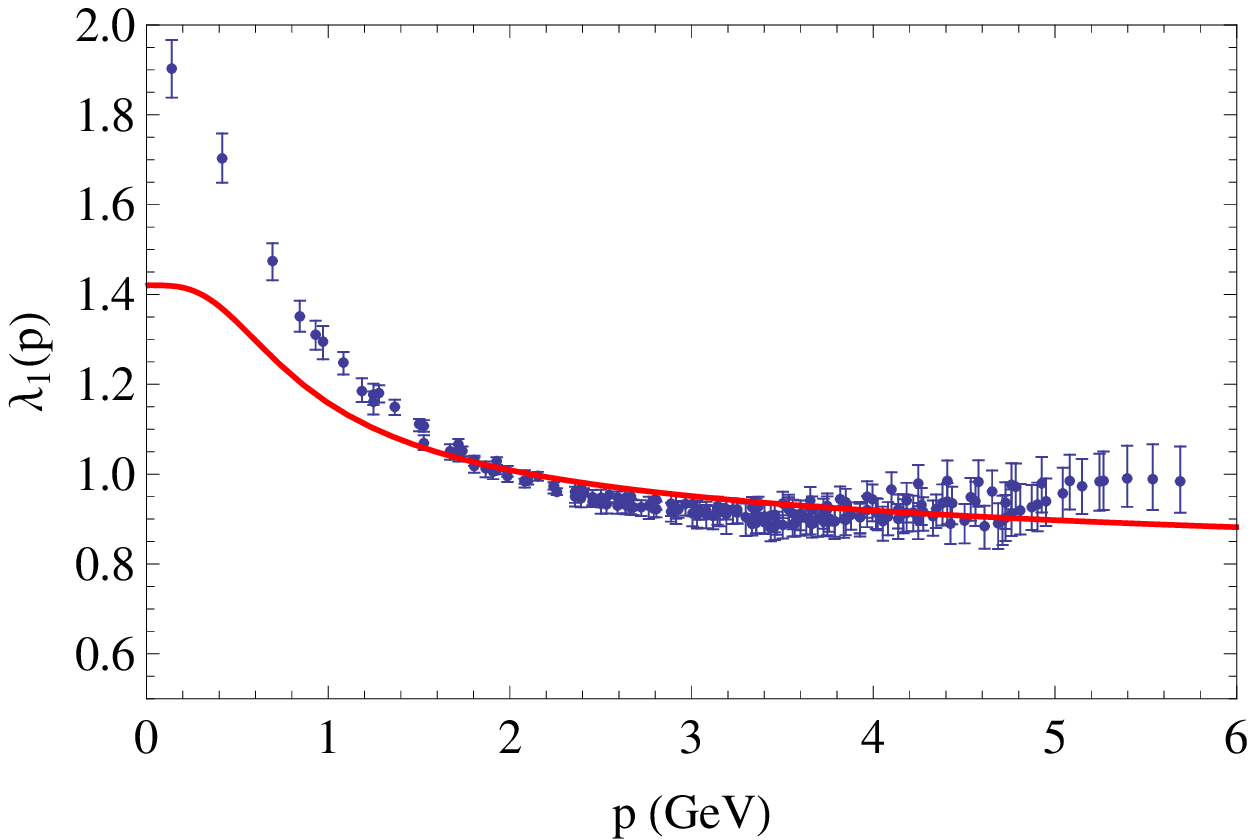}
 \includegraphics[width=\linewidth]{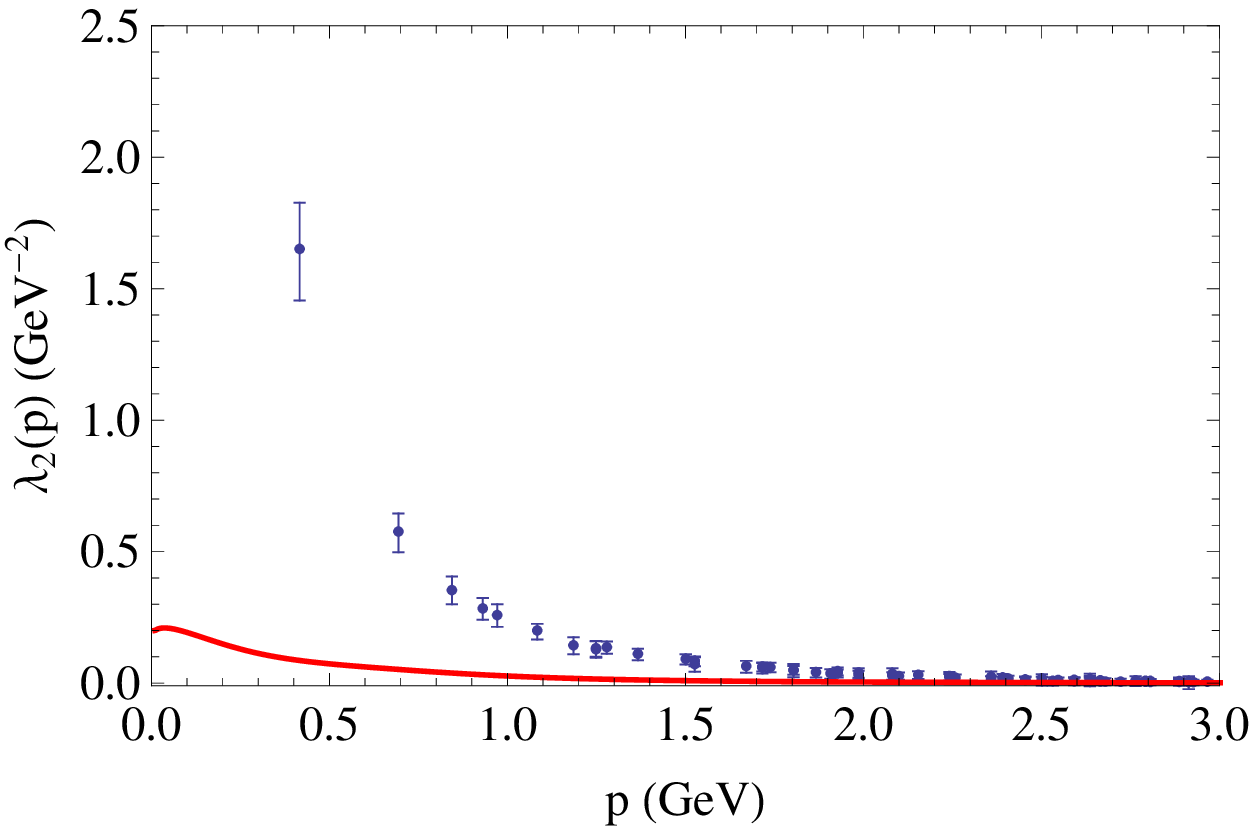}
 \includegraphics[width=\linewidth]{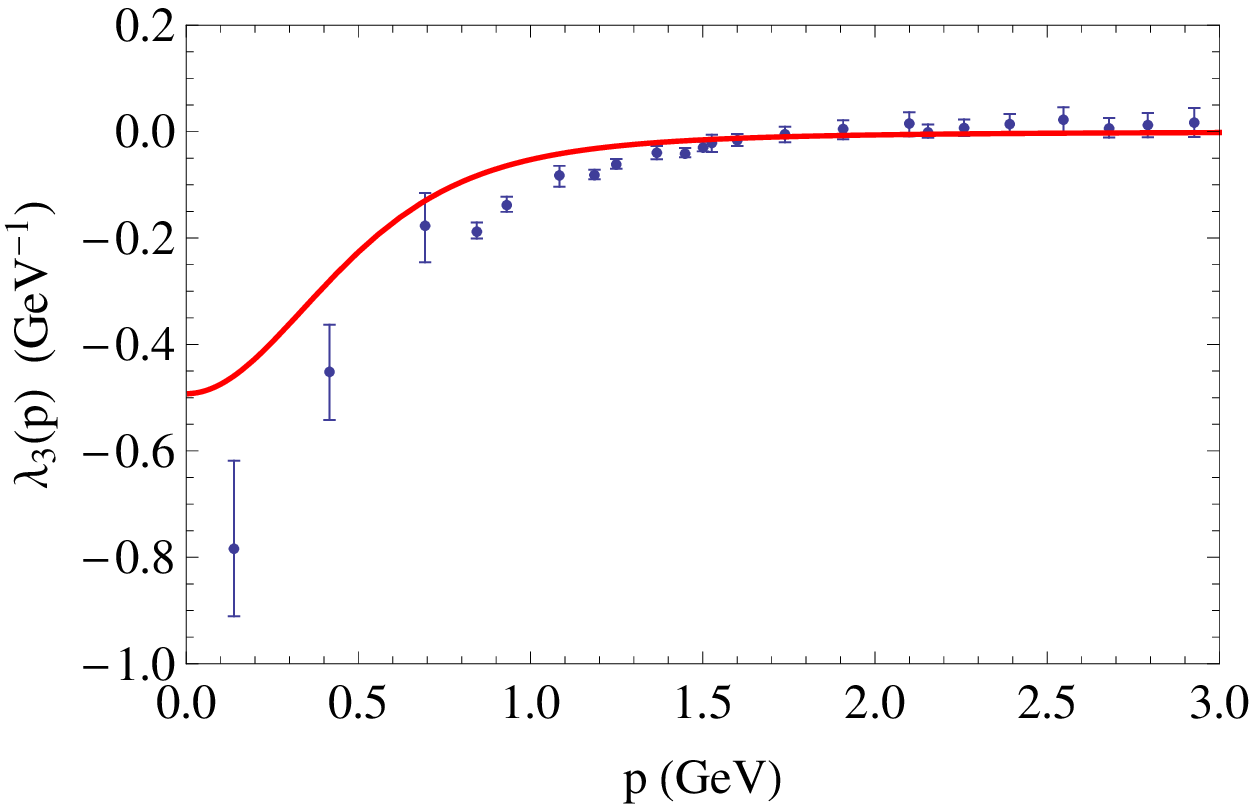}
 \caption{\label{FunEsck0} Scalar functions of the quark-gluon vertex:
   $\lambda_1$ (top), $\lambda_2$ (middle) and $\lambda_3$ (bottom)
   for a vanishing gluon momentum, as a function of the quark
   momentum. The line are the one-loop results with $M=0.2$~GeV.
   The dots correspond to the lattice data of
   \cite{Skullerud:2003qu}.}
\end{figure}

The resulting curves for the functions $\lambda_1(p^2)$,
$\lambda_2(p^2)$ and $\lambda_3(p^2)$ can bee seen in
Fig.~\ref{FunEsck0}.
{ We find a good agreement for the function $\lambda_1(p^2)$. All the point are
below a 16\% of error except the first point ($p=140$~MeV) where a larger
error is observed (27\%). Let us point out, however that our curve is systematically below
the lattice one. The behavior for $\lambda_3(p^2)$ is qualitatively
reproduced, with a strong increase (in absolute value) at momenta of
the order of 1~GeV This last curve depends more strongly on the quark
mass. This can be understood by the observation that $\lambda_3$ tends
to zero in the chiral limit, see Table~\ref{tab_struc}. The function is very well
reproduced (essentially within error bars) except for the two lowest momenta
($p=$~140 and 400 MeV) where the errors are larger (37\% and 33\% respectively when comparing to the lattice central value).
Let us, however, point out that the statistical error bars of these lattice points are large
(22\% in both points).

We therefore find that for a unique
choice of parameters $g$, $m$ and $M$, we reproduce very well the ghost and
gluon propagators and reasonably well the mass function $M(p)$ and the vertex functions $\lambda_1$ and $\lambda_3$,
which is a highly nontrivial result.}

The situation looks qualitatively different for the function
$\lambda_2(p^2)$, which tends to a constant at zero momentum in our
approach while it diverges in the lattice data. This disagreement was
already observed in previous analytical treatments of this function
\cite{Williams:2014iea,Aguilar:2014lha} and was interpreted as a
consequence of the non-perturbative features of the theory. We want to
propose a completely different explanation for this phenomenon. Let us
go back to Eq.~(\ref{eqL2}) which is used to extract $\lambda_2$ from
the lattice data. We observe that we must subtract from the function
$\lambda_1(p^2)$ a quantity
\begin{equation}
  \label{eq_def_lambda2t}
\widetilde  \lambda_2=-\frac{1}{16g_B}\sum_{\mu}\text{Im}\,\text{Tr}\, \left[\gamma_\mu\Gamma_\mu(p,-p,0)\right].
\end{equation}
which is directly extracted from lattice data.  The difference is then
divided by $p^2$. In our calculation, the fact that $\lambda_2$ tends
to a constant originates from a compensation of these two terms in the
limit $p\to 0$, the difference being of order $p^2$. We can interpret
this constant as a compensation of large numbers, a situation which is
difficult to treat numerically. Indeed, a small error in one of these
two terms would lead to a divergence of $\lambda_2(p^2)$ when $p\to 0$
but this would be certainly an artifact.

Note that such a divergence in $\lambda_2(p^2)$ would imply a
nonanalytic behavior of the vertex function at small momenta [see
Eq.~(\ref{zerogluonmomentum})] since the $p\to 0$ limit would depend
on the direction (in momentum space) which is used to perform this
limit. In our calculations, which are done in the framework of the
action (\ref{eq_cf})--(\ref{eq_quarks}), the only non-analytic behaviors
observed so far are a consequence of ghost loops (see, for example,
\cite{Tissier:2010ts,Tissier:2011ey,Pelaez:2013cpa}). Such
contributions are not present at one-loop order in the present vertex
and most probably their contributions of higher order should be
extremely small.

\begin{figure}[tbp]
\centering
 \includegraphics[width=\linewidth]{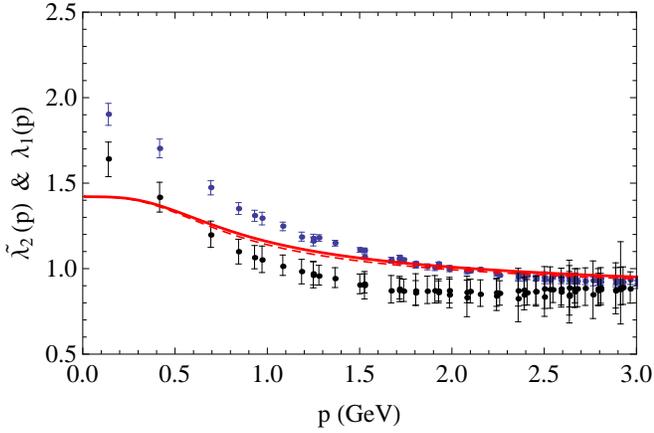}
\caption{\label{FunEsck0L1L2comb} { Comparison of the quantities
  $\lambda_1$ and $\tilde \lambda_2$ as a function of the momentum
  [see Eq.~\ref{eq_def_lambda2t}].  Our one-loop results for
  $\lambda_1$ (full line) are almost degenerate with those for $\tilde
  \lambda_2$ (dashed line). The higher (blue) points correspond to
  $\lambda_1$ and the lower (grey) ones to $\tilde\lambda_2$,
  extracted from \cite{Skullerud:2003qu}. In the latter case, our
  estimation of the error bar is probably pessimistic.} } 
\end{figure}

If the disagreement takes it origin in partial compensations in the
extraction of the $\lambda_2(p^2)$ function from the formula
(\ref{eqL2}), the comparison should be better for the quantity that is
directly extracted from lattice data (that is $\widetilde \lambda_2$)
than that for $\lambda_2$ itself. In Fig.~\ref{FunEsck0L1L2comb} we
compare our prediction for $\widetilde \lambda_2$ to the corresponding
lattice values. We observe a very good agreement in both expressions:
{our curve is almost within the lattice error
  bars except for the lowest momentum where the error is around the 17\% with
  respect to the central value of simulations. We must, however point out that the
  curve is systematically above the central values of lattice data.
  The agreement between the $\widetilde \lambda_2$ with lattice data suggests that the true origin of the disagreement
between analytical and lattice data for $\lambda_2$ lies in the difficulty of
obtaining a function from differences of large numbers.} Let us point
out that the origin of the error is certainly not statistical. If it
were so, the signal obtained from the lattice data for
$\lambda_2(p^2)$ would be extremely noisy. Most probably the origin of
the error is more systematical. Now it is important to recall that the
lattice extraction of these scalar function requires a careful
treatment of various lattice artifacts (see
\cite{Skullerud:2002ge,Skullerud:2003qu}). A mismatch in the treatment
of lattice artifact of the longitudinal and transverse parts of the
vertex could easily introduce a systematic error that, when divided by
$p^2$, gives a large spurious contribution in the estimate of
$\lambda_2(p^2)$. In any case, Fig.~\ref{FunEsck0L1L2comb} shows
clearly that not only the component transverse to the external
momentum but also the longitudinal one are correctly reproduced by the
present one-loop calculation.

\subsection{Quark and anti-quark with equal momenta}

In this section we  present the results for the kinematic
configuration given by equal momenta of quark and anti-quark ($r=p$)
and, accordingly, $k=-2p$.  In this kinematics the quark-gluon vertex
simplifies considerably
\begin{align*}
\Gamma_\mu(p,p,-2p)&= -ig\left[\lambda_1(p^2)\gamma_\mu+4\tau_3(p^2)(\slashed{p}p_\mu-p^2\gamma_\mu)\right.\nonumber\\
&\left.-2i\tau_5(p^2)\sigma_{\mu\nu}p_\nu \right]
\end{align*}
and the transverse projected vertex (that is
$P^\perp_{\mu\nu}(p)\Gamma_\nu(p,p,-2p)$)< has the form
\begin{equation*}
\Gamma_\mu^P(p,p,-2p)=-ig\left[\lambda'_1(p^2)\left(\gamma_\mu-\frac{\slashed{p}p_\mu}{p^2}\right)-2i\tau_5(p^2)\sigma_{\mu\nu}p_\nu\right]
\end{equation*}
where we have introduced $\lambda_1'=\lambda_1-k^2\tau_3$.

In spite of this simplification, for this configuration of momenta,
integrals can not be done analytically and some remaining Feynman
parameter integrals must be performed numerically. Fortunately, having
fixed the parameters { from propagators}, the
calculation must be done only once for a given choice of parameters.

In this kinematic configuration, the scalar functions $\lambda'_1$ and
$\tau_5$ are extracted from the vertex as
\[\lambda'_1=-\frac{1}{3}\sum_{\mu}\frac{1}{4g_B}\text{Im}\left(\text{Tr}\, \gamma_\mu\Gamma_\mu^P(p,p,-2p)\right)\]
and
\[
\tau_5=\frac{1}{3k^2}\sum_{\mu,\nu}k_\mu\frac{1}{4g_B}\text{Re}\left(\text{Tr}\,
  \sigma_{\mu\nu}\Gamma_\mu^P(p,p,-2p)\right).\] In this case, no
subtraction is needed to extract $\lambda'_1$ and $\tau_5$ from the
lattice data.

\begin{figure}[tbp]
\centering
 \includegraphics[width=\linewidth]{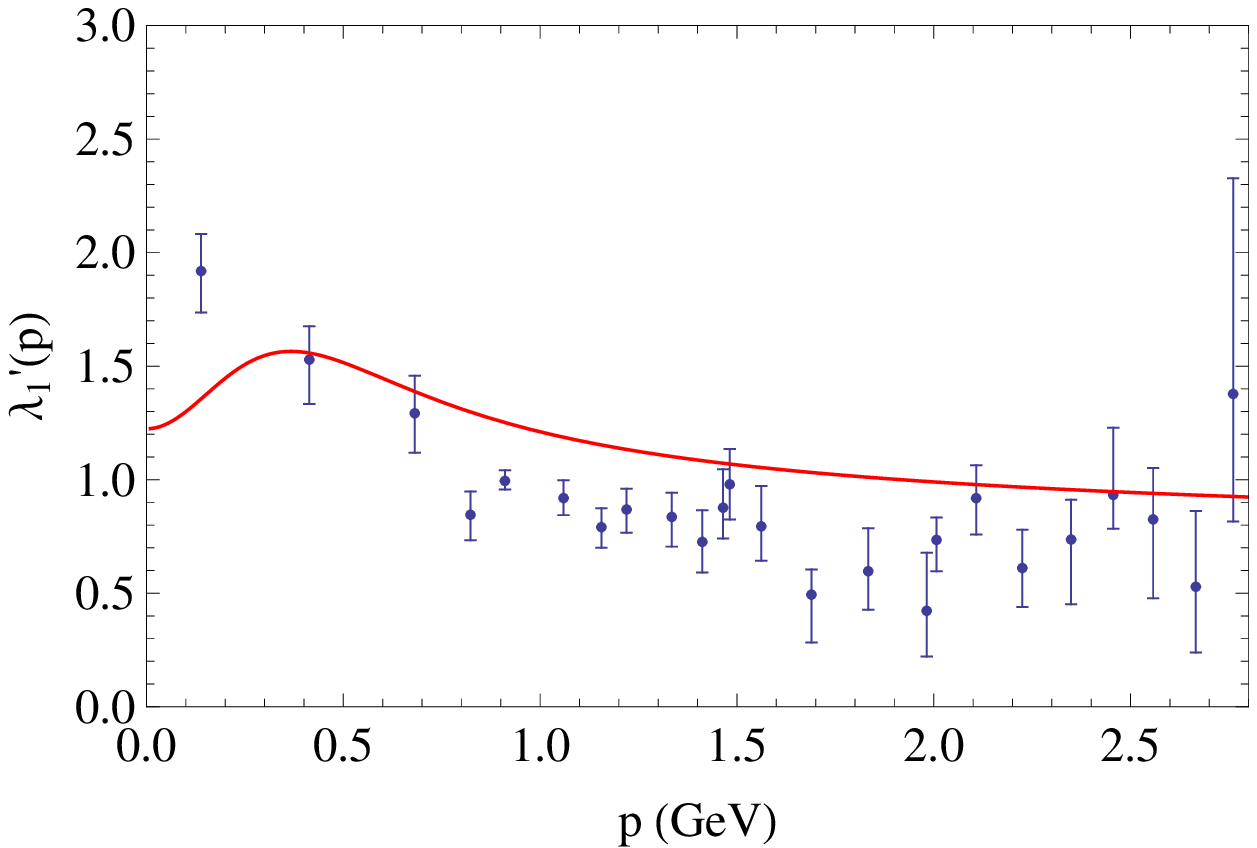}
 \includegraphics[width=\linewidth]{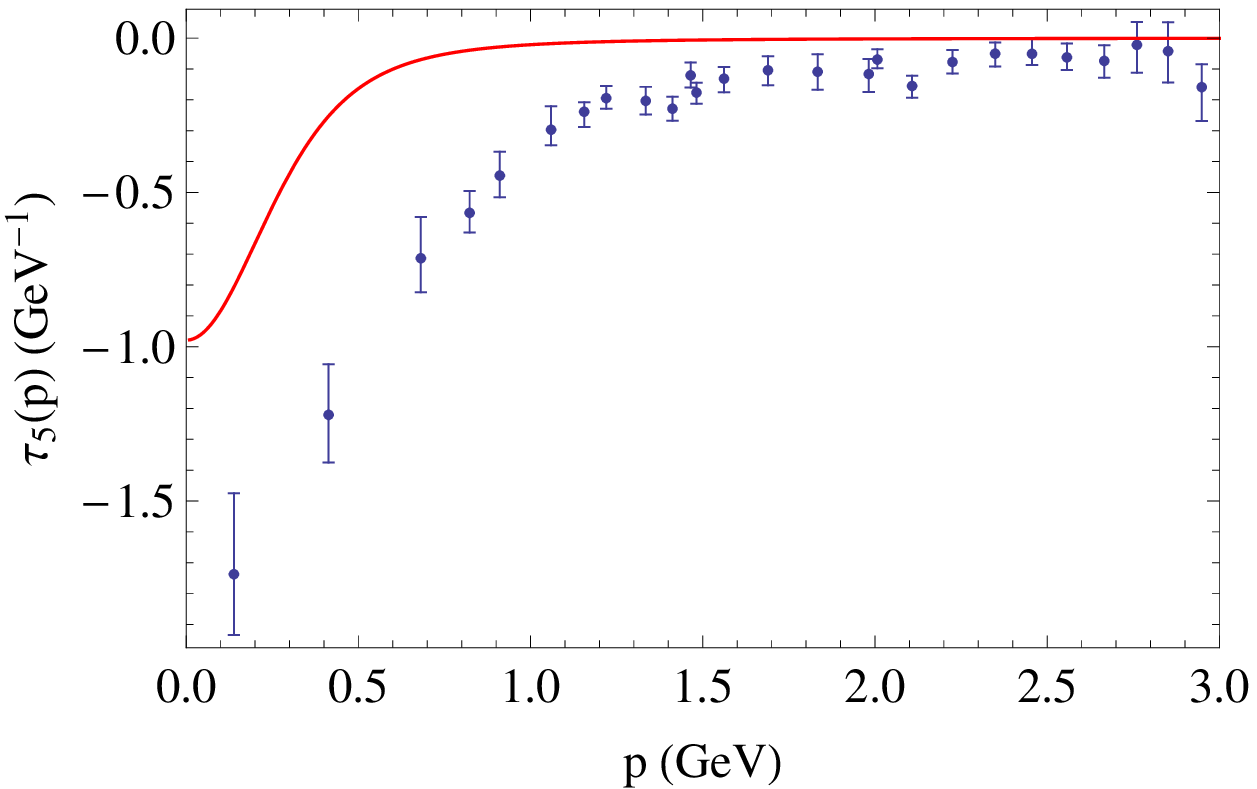}
 \caption{\label{FunEssim} Scalar function of the quark-gluon vertex:
   $\lambda_1'$ (top) and $\tau_5$ (bottom) for equal momentum (p) for
   the quark and antiquark. {The full line are the one-loop results
   and the dots correspond to the lattice data
   of~\cite{Skullerud:2003qu}.}}
\end{figure}

{In Fig.~\ref{FunEssim} we compare our one-loop results with the
lattice data using the same initial condition of the coupling constant
and masses as in the previous case, $g=4.2$, $m=0.44$ GeV and for
the value of $M=0.2$ GeV at $\mu_0=1$ GeV.  We observe that the function $\lambda_1'$ is well
reproduced by our results (mostly within the lattice
  error bars, except for the lowest momentum point where the error is
  of the order of 25$\%$). The other function $\tau_5$} is seen to be very
sensitive to the choice of the quark mass at 1 GeV. Again, this is a
consequence of the fact that, in the chiral limit, this function tends
to zero, see Table~\ref{tab_struc}. We observe that the qualitative
behavior for this function is reproduced but the agreement is not
quantitative. In particular, our one-loop results do not reach such
large values in the deep infrared { where for the lowest
value of momenta the error reach 45\%. Moreover, the curve is systematically
above the lattice curve.} This is similar to what has already
been observed in the DS results of \cite{Aguilar:2014lha}.

\subsection{Symmetric configuration}

In this subsection we consider the case where the quark and antiquark
have momenta with the same modulus but can have arbitrary angle
between them (or, equivalently, that one can vary the value of the
gluon momentum). It is clear that the two previous kinematic
configurations correspond to particular cases of the more general
situation considered here. However, more general configurations of
momenta were studied in the lattice \cite{Kizilersu:2006et}, but only
for a particular tensorial structure.  More precisely, in terms of the
tensorial decomposition (\ref{eq:decomp_2}), the function that is
extracted in the lattice is $\lambda_1'=\lambda_1-k^2\tau_3$ as in the
previous case. We limit to three particular kinematic
configurations. In the first two, the modulus of the gluon momentum
taking values $|k|=0.277$ GeV and $|k|=0.838$ GeV. The third
corresponds to the completely symmetric case where the modulus of the
three momenta are equal $r^2=p^2=k^2$. { As before, initial conditions
of the renormalization group equations correspond again to $g=4.2$,
$m=0.44$ GeV and $M=0.2$ GeV at $\mu_0=1$ GeV}.

\begin{figure}[tbp]
\centering
 \includegraphics[width=\linewidth]{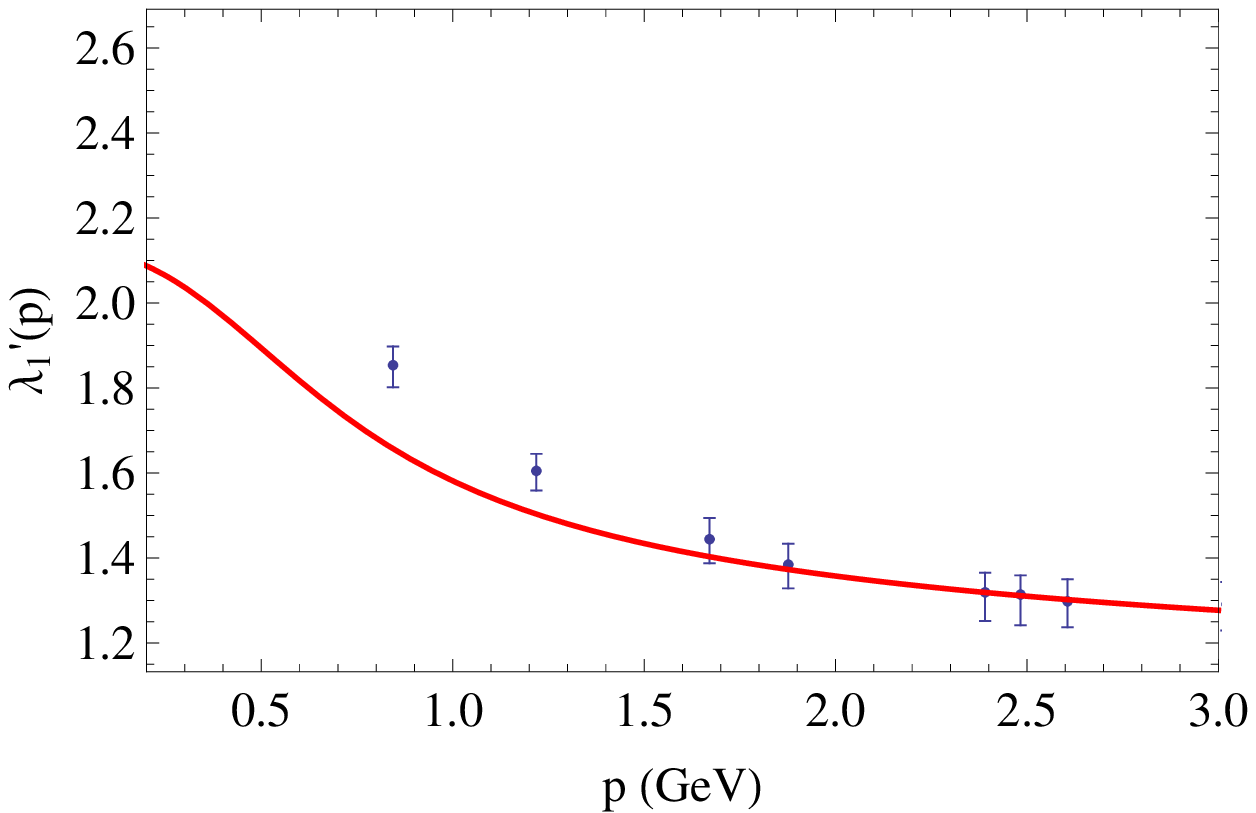}
 \includegraphics[width=\linewidth]{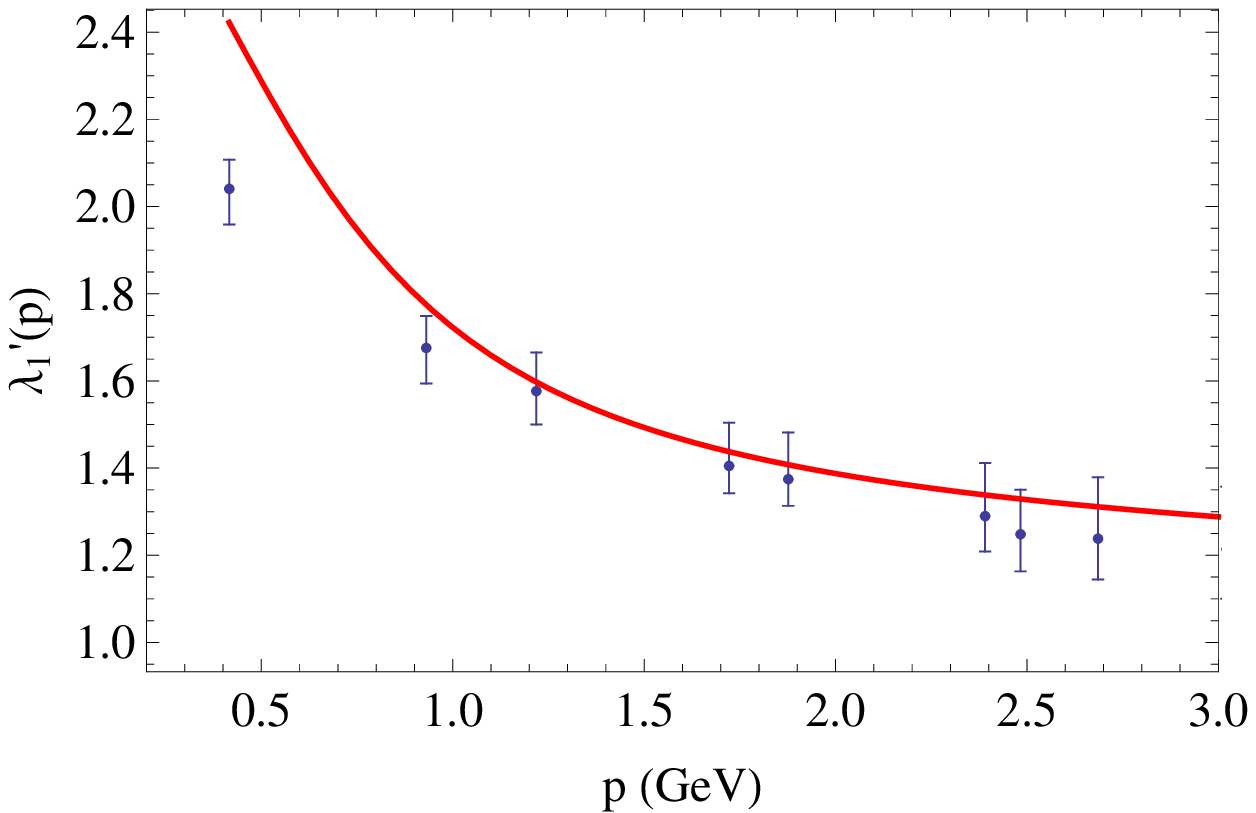}
 \caption{\label{Fun_fully_sym} Scalar function of the quark-gluon
   vertex $\lambda_1'=\lambda_1-k^2\tau_3$ for $r^2=p^2$ and $k=0.277$
   GeV (top) and $k=0.838$ GeV (bottom). The full line are the one
   loop results and the dots the lattice data of
   \cite{Skullerud:2004gp}.}
\end{figure}

\begin{figure}[tbp]
\centering
 \includegraphics[width=\linewidth]{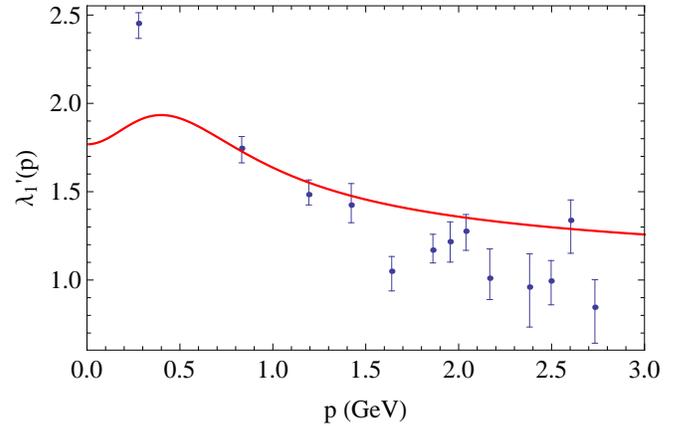}
\caption{\label{Fun_fully_sym_2} Scalar function of the quark-gluon vertex $\lambda_1'=\lambda_1-k^2\tau_3$
for the completely symmetric configuration $r^2=p^2=k^2$. The full line are the one loop results and the dots the lattice data of \cite{Skullerud:2004gp}
interpolated to reconstruct the symmetric configuration.} 
\end{figure}

{The results are presented in Figs.~\ref{Fun_fully_sym} and
\ref{Fun_fully_sym_2}. For $|k|=0.277$~GeV the maximal error is of
12$\%$. For $|k|=0.838$~GeV, our results are within lattice error bars
with the exception of the point of lowest momentum ($p = 0.4$ GeV)
where the error is of 18$\%$. In the completely symmetric configuration
the curve is almost within error bars except for the lowest momentum
where the error is 25\%.} At the end of this section we discuss on
possible origins of these discrepancies.

\subsection{Unquenched vertex}
\label{sec:unquenched}

We conclude our study by introducing the effect of dynamic quarks in
our calculations. This influences the renormalization factor for the
gluon $Z_A$ and for the gluon mass $Z_{m^2}$, which both have a
contribution from a fermion closed loop. On the contrary, the first
contribution of dynamic quarks to $Z_c$ and to the quark-gluon vertex
comes at two loops. We conclude that the effect of dynamic quarks only
influence indirectly our calculation, through the modification of the
$\beta$ functions for the coupling constant and the gluon mass and for
the field renormalization $z_A$.

The unquenched propagators for the quarks, gluons and ghosts where
studied in \cite{Pelaez:2014mxa}, in the framework of the
Curci-Ferrari action. { We use here the parameters that were obtained in
that article, as best fits when comparing with the lattice data in
the only case where the mass quark can be fixed from the a quark mass function, that is
$N_f=2+1$. The best fit parameters in this case is $g=4.8$, $m=0.42$~GeV and
$M=0.08$~GeV at $\mu_0=1$~GeV.}
The corresponding values of the couplings were compared
in \cite{Pelaez:2014mxa} with standard estimates of the coupling giving
an agreement with an error of the order of 17 \%, which is similar to the typical overall
error in the infrared of the one-loop approximation.

Using these values of the parameter we can now predict the vertex in
the unquenched case. It is interesting to note that in the present
procedure this is an extremely simple calculation once the beta
functions have been calculated and the parameters fixed by comparing
to the 2-point functions. This is in contrast to lattice simulations
that are extremely costly in the unquenched case. This is probably the
reason why there are, to our knowledge, no lattice data for this
vertex. 

\begin{figure}[tbp]
\centering
 \includegraphics[width=\linewidth]{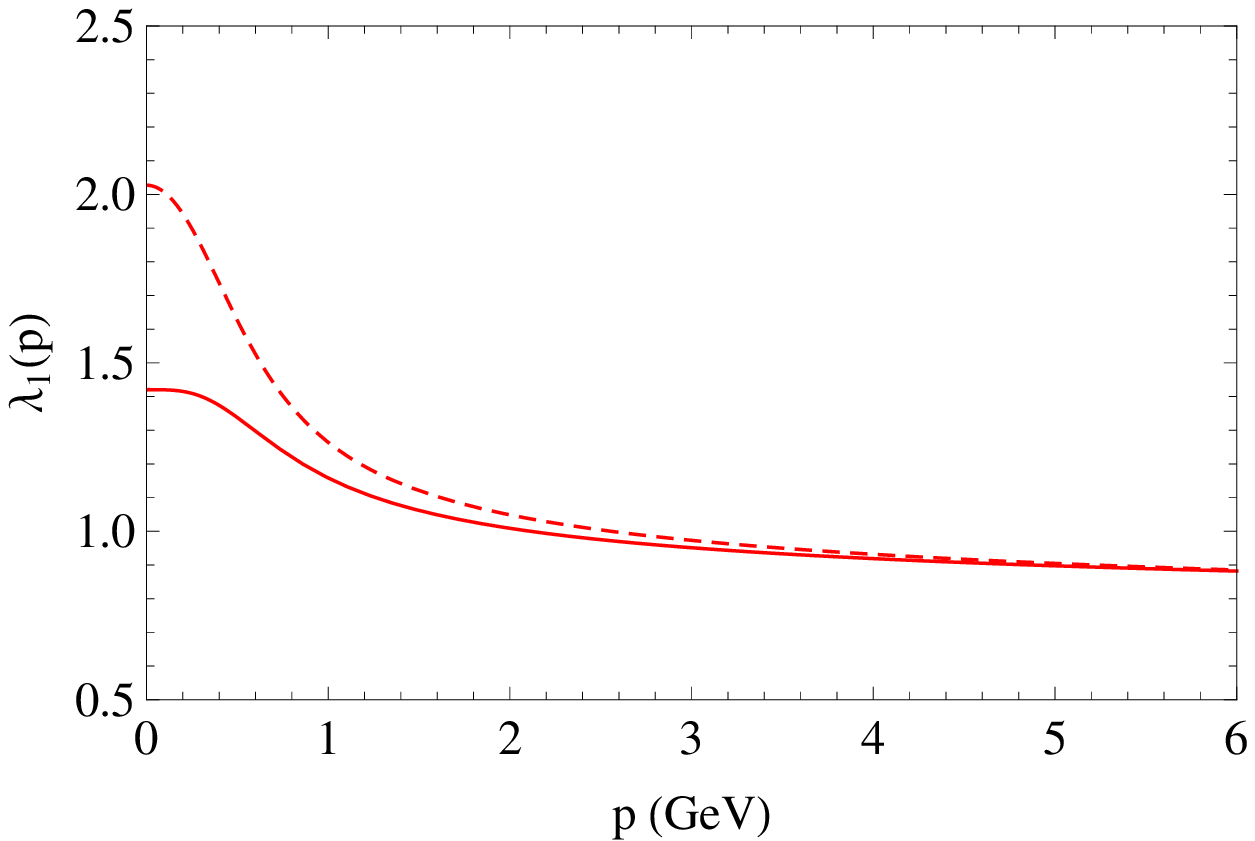}
 \includegraphics[width=\linewidth]{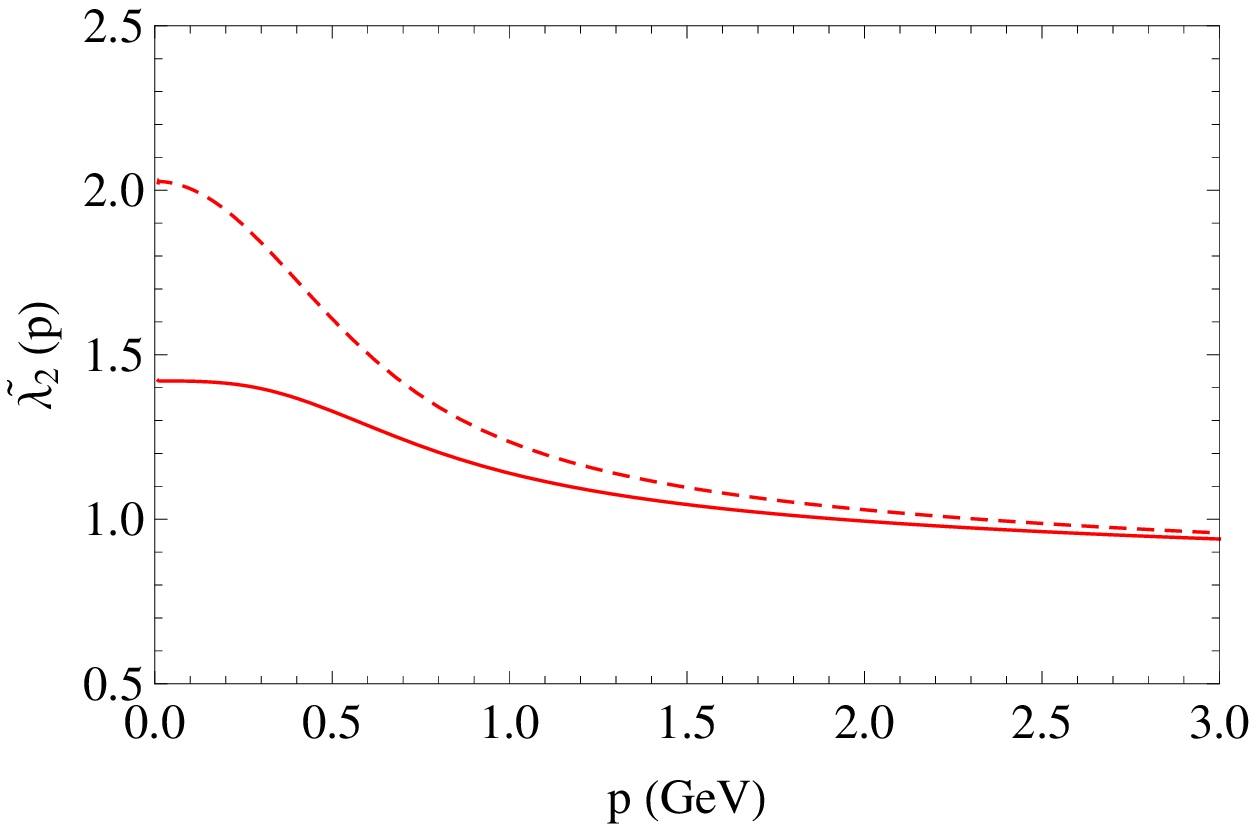}
 \includegraphics[width=\linewidth]{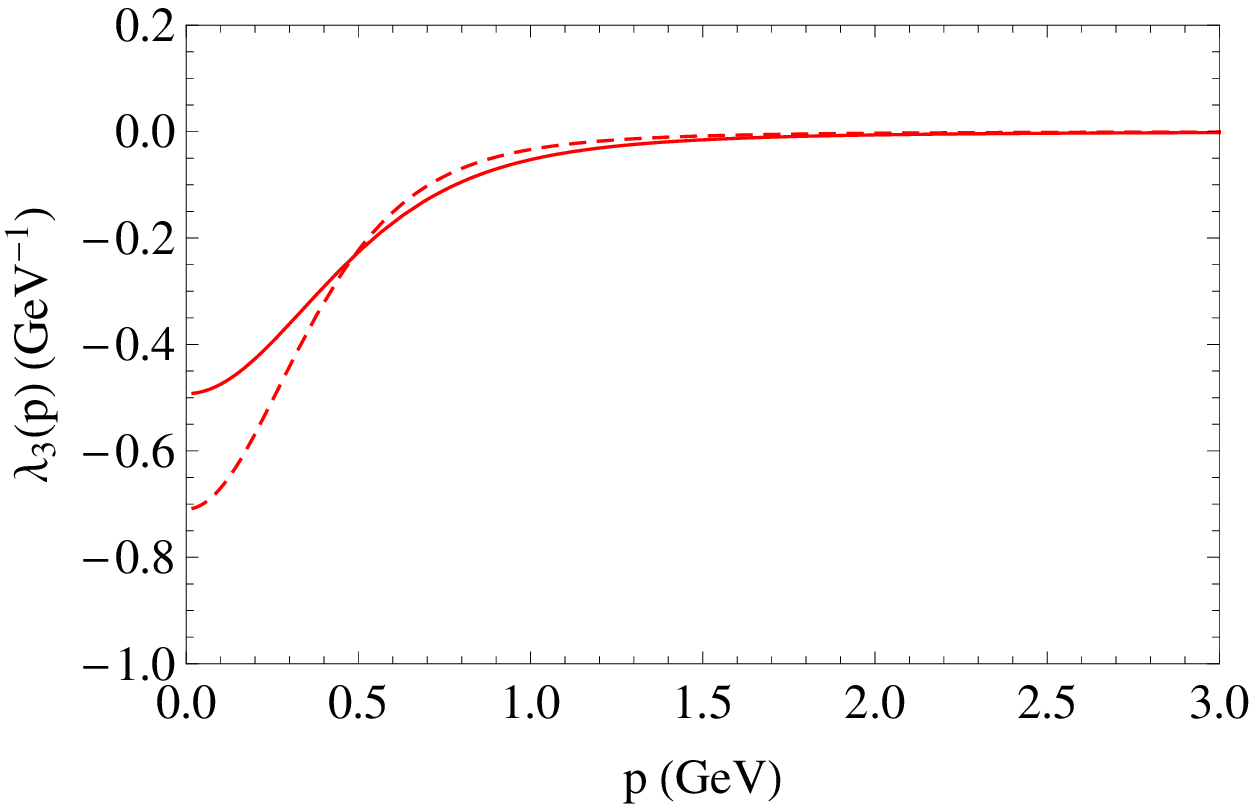}
 \caption{\label{Unquenchedcurves} Scalar functions of the quark-gluon
   vertex in the unquenched case.  $\lambda_1$ (top), $\tilde
   \lambda_2$ (middle) and $\lambda_3$ (bottom) for a vanishing gluon
   momentum, as a function of the quark momentum. The curves corresponds to For $N_f=0$ (full
   line) and $N_f$=2+1 (small dashes).}
\end{figure}

We focus here on the zero gluon momentum but other tensorial
structures and kinematic configuration could be presented as well. We
present in Fig.~\ref{Unquenchedcurves} the three scalar functions
$\lambda_1$, $\tilde\lambda_2$ and $\lambda_3$ for different number of
flavors. The presence of dynamic quarks clearly tends to increase
these functions at low energy.

\subsection{Possible origin of discrepancies}

The overall agreement of the present one-loop calculation is { reasonably good}.
Most of the lattice data are well reproduced with a good
accuracy. It is important to stress that most of the calculations are
pure predictions without free parameters (that are already adjusted
via 2-point functions) and that they are all obtained by one loop
calculations.

However, some discrepancies are present and we analyze them in the present subsection.
Putting aside the $\lambda_2(p)$, the largest discrepancy
concerns the $\tau_5(p)$ function in the equal momenta
configuration. Even if in this case the order of magnitude of the
function and general behavior is well reproduced, there is not
quantitative agreement.  Finally, one observes some quantitative
disagreement for the function $\lambda_1'(p)$ function in the
symmetric configuration at small momenta. We can say that there is a good
general agreement with lattice data but the quantitative agreement is
not as good as for gluon and ghost vertices. A similar behavior is
observed in the study of propagators \cite{Pelaez:2014mxa}.

We give now two possible (probably complementary) explanations for
this. First, one of the reasons of the success of a perturbative
analysis in the gluon and ghost sector is that the relevant couplings
(the value of the 3-gluon and ghost-gluon couplings) are moderate
except, for the 3-gluon vertex for very small values of the coupling
where the presence of the gluon mass, has already suppressed
fluctuations. However, in the quark sector, { even if the coupling does not
presents a Landau pole, it reaches significantly larger values than in the pure-glue sector.}
One measure of this is the value of the coupling $\lambda_1(p)$
or $\lambda_1'(p)$. One can define a natural running coupling function
for this functions as
$\alpha_s^{\text{quark-gluon}}(p)=(\lambda_1'(p))^2
\alpha_s^{\text{ghost-gluon}}(p)$. It is important to stress that even
if for $p\gg m$, all the definitions of the running coupling give a
universal beta function, this is not the case for momenta $p\lesssim
m$. In consequence, the moderate values of the coupling in a certain
scheme does not exclude that it may be large in some other scheme. In
particular, one observes that for small values of $p$ the quark-gluon
coupling is larger by a factor of 2.4 with respect to the ghost-gluon
coupling (giving a larger expansion parameter by a factor
$(2.4)^2\simeq 5.8$. In a sense, this is to be expected because it is
well established (see, for example, \cite{Alkofer:2008tt}) that the
quark-gluon coupling that is required for generating Dynamical
Chiral Symmetry Breaking (DSCB) is larger that the one that is
extracted from the ghost-gluon vertex. Now, the expansion parameter of
the perturbative expansion in the present model in the ghost-gluon
sector (precisely discussed in \cite{Tissier:2011ey}) may reach values
of the order of 0.2. In consequence, the expansion parameter in the
quark-gluon sector seem to put the perturbative expansion in trouble
at very low momenta.  This may indicate that even in the present model
where the ghost-gluon sector seem to be treatable perturbatively, some
sort of partial resummation is needed in the quark sector (or, at
least, in part of it). { These facts also reflects on a certain tension
when the parameters have to be chosen in order to fit all the data at the same time.
Even if the overall agreement is good with a single set of parameters, a precise
renormalization scheme had to be chosen because at very low momenta an important
sensitivity is observed.}

This last point is related to a second possible origin of
discrepancies. In the quark-gluon vertex, the quantities
that are not sensitive to the DCSB are much better es-
timated that those that are sensitive to it. The function $\lambda_1(p)$, gives better agreement than
the $\lambda_3(p)$ or the $\tau_5(p)$ (which are sensitive to the
DSBC). This manifests also in the fact that those quantities seem to
have a larger sensitivity on the renormalization conditions. Now, it
is not clear in what measure we have properly taken into account the
role of the DSBC. Even if RG effect gives an important enhancement of
the quark mass in the infrared that quantitatively reasonably
reproduces the mass function measured in the lattice, it is unclear at
the moment if a proper treatment DSBC will require to use more
sophisticated techniques as the introduction of composite fields for
the chiral condensate. This may be at the origin of the lower quality
of quantities that are sensitive to the DSBC.

\section{Conclusion}

In this article, we have presented the calculation of the quark-gluon
vertex at one loop in the massive (Curci-Ferrari) extension of the
Landau gauge. The results were decomposed on the twelve independent
tensorial structures. We implement the renormalization-group
improvement so as to avoid large logarithms. We find that these
analytic results compare qualitatively well with the quenched lattice
data. In one case ($\lambda_2$ for a vanishing gluon mass), the
difference between lattice data (that predicts a divergence at low
momentum) and our (which are finite at small momenta) may be
attributed to a difference of large numbers when extracting this
function from lattice data.

At the quantitative level, we observe that the quantities that are not
forbidden by chiral symmetry are reproduced at the level of 10\% {
except for momenta of the order of 100-200 MeV where larger errors are found,
as detailed in the results section}. The
others are much more difficult to reproduce and show strong dependence
on the renormalization-group scheme. This is probably to be attributed
to the rather crude treatment of chiral symmetry breaking in our
calculation. { Moreover, in order to fit simultaneously
both the mass function coming from the quark propagator and the vertex function,
a certain tension in the choose of parameters is observed and in order for
this to be fulfilled it is necessary to choose a precise renormalization scheme.}
It is interesting to point out that even if{, at the end,} the agreement
of the present one-loop calculation is reasonably good, the precision
is not as good as in the ghost-gluon sector. This may be related to a
larger value of the associated coupling and some sort of resummation
in this sector could be useful. This issues will be addressed in the
future.

\begin{acknowledgments}
  We thank J.-I.~Skullerud for kindly making available the lattice
  data and for useful exchanges.  We acknowledge partial support from
  PEDECIBA and ECOS-Sud (U11E01) programs.
\end{acknowledgments}

\end{document}